\documentclass[a4paper,11pt]{article}
\pdfoutput=1 % if your are submitting a pdflatex (i.e. if you have
\usepackage{jheppub} % for details on the use of the package, please
\usepackage[T1]{fontenc} % if needed
\usepackage{subfig}
\usepackage{slashed}
\usepackage{mathtools}
\usepackage{enumitem}
\usepackage{braket}
\usepackage{xfrac}
\usepackage[capitalise]{cleveref}

\def\d{\partial}
\def\nl{\nonumber \\}
\def\nquad{\mkern-24mu}

\usepackage[normalem]{ulem}

\title{\boldmath Dark matter in the Randall-Sundrum model with non-universal coupling}

\author[a,b]{Ashok Goyal,}
\author[a,c]{Rashidul Islam,}
\author[a,1]{and Mukesh Kumar\note{Corresponding author.}}

\affiliation[a]{School of Physics and Institute for Collider Particle Physics, 
University of the Witwatersrand, Johannesburg, Wits 2050, South Africa.}
\affiliation[b]{Department of Physics and Astrophysics, University of Delhi, New Delhi, Delhi, India.}
\affiliation[c]{Department of Physics, Indian Institute of Technology, Guwahati, Assam 781039, India.}

\emailAdd{agoyal45@yahoo.com}
\emailAdd{rislam@iitg.ac.in}
\emailAdd{mukesh.kumar@cern.ch}

\abstract{We consider simplified dark matter models (DM) interacting gravitationally with the standard model (SM) particles in a Randall-Sundrum (RS) framework. In this framework, the DM particles interact through the exchange of spin-2 Kaluza-Klein (KK) gravitons in the $s$-channel with the SM particles. The parameter space of the RS model with universal couplings to SM particles is known to be strongly constrained from the LHC data. We are thus led to consider models with non-universal couplings. The first model we consider in this study is a top-philic graviton model in which only the right-handed top quarks are taken to interact strongly with the gravitons. In the second, the lepto-philic model, we assume that only the right-handed charged leptons interact strongly with the gravitons. We extend the study to include not only the scalar, vector and spin-1/2 fermions but also spin-3/2 fermionic dark matter. We find that there is a large parameter space in these benchmark models where it is possible to achieve the observed relic density consistent with the direct and indirect searches and yet not to be constrained from the LHC data.}

\begin{document} 
\maketitle
\flushbottom

\section{Introduction}
\label{sec:intro}
Dark matter (DM) existence has been inferred from several cosmological and astrophysical observations at different scales. At the galactic scale, we have the observation of the flattening of rotation curves, weak lensing measurements at the scale of galactic clusters and the CMB and large scale structure observations at the cosmic scale point towards the existence of DM in the Universe. The dark matter constitutes roughly 75\% of the entire matter existing in the Universe. The Planck collaboration has measured the DM density to great precision and has given the value of relic density $\Omega_{\rm DM} h^2 = 0.1198\pm0.0012$~\cite{Aghanim:2018eyx}. ($\Omega_{\rm DM}$ is the DM mass density in units of critical density and $h \sim 0.7$ is today's Hubble constant in the units of $100~{\rm km/s/Mpc}$.) The nature of the DM particles, however, remains elusive. One of the ways to unravel the nature of DM particles is to search for its non-gravitational interaction with the Standard Model (SM) particles, but none has been manifested so far. Direct searches of DM particles from the local galactic halo aim to measure recoil of nucleons through scattering in underground targets. In direct detection, the important quantity is the spin-independent (SI) and spin-dependent (SD) scattering cross-sections in a non-relativistic (NR) regime. These measurements have reached a sensitivity level where $\sigma_{\rm SI} > 8\times 10^{-47} {\rm cm}^2$ for DM masses $\sim 30$~GeV has been ruled out in Dark Side~50 (2016)~\cite{Agnes:2015ftt}, LUX~\cite{Akerib:2016vxi}, XENON~\cite{Aprile:2017iyp}, and PANDA~\cite{Cui:2017nnn}. Indirect detection aims at unveiling excess cosmic rays produced by DM annihilation or decay in the galaxy or beyond. The indirect detection of DM is made through the emission of monochromatic gamma rays by the satellite-based $\gamma$-ray observatory Fermi-LAT~\cite{Fermi-LAT:2016uux,Ackermann:2015lka} and ground-based Cherenkov telescope H.E.S.S~\cite{Abramowski:2013ax}. Collider searches by the ATLAS and CMS collaborations~\cite{Garny:2013ama, Goodman:2010yf} at the Large Hadron Collider (LHC) aim at identifying signatures of production of DM particle involving missing energy (${\slashed E}_T$) accompanied by a single (monojet +${\slashed E}_T$) or two jets (dijet + ${\slashed E}_T$) events.

No experimental observation so far has made any confirmed detection and as a result, a large DM parameter space has been excluded. Most of the theoretical effort has been invested as the hypothesis that DM is a weakly interacting massive particle (WIMP) with mass lying between several GeV to a few TeV. WIMPS emerge in attempts to address the hierarchy problem underlying the staggering difference between the Planck Scale ${\cal O}(10^{16}$~GeV) on the one hand and the electro-weak scale ${\cal O}(100$~GeV) on the other. WIMPS provide the simplest production mechanism for massive relics from the early Universe. Freeze-out from the thermal plasma for particles having weak scale interactions and mass lying between GeV to TeV range can naturally account for the observed relic density. The fact that recent detection experiments notably by XENON1T have not found a signal puts a severe strain on the parameter space of one of the most attractive theories to address the hierarchy problem namely, the supersymmetric (SUSY) theories and, most of the parameter space of natural simple SUSY is ruled out~\cite{Athron:2017qdc}. Null results are constraining more and more of the parameter space of current DM theories.

Randall-Sundrum (RS) formalism in the 5D warped extra-dimensions~\cite{Randall:1999ee} with its Kaluza-Klein (KK) gravitons interacting purely gravitationally with the SM particles provide another possible solution to the hierarchy problem. The RS model with universal coupling to all SM particles is, however, very seriously constrained in its parameter space by the collider (LHC) data~\cite{Kraml:2017atm} and one is thus led to consider the bulk RS model with non-universal couplings. In one of these models, KK graviton interacts strongly with only the top-quark and this making the top-quark loops important for its production and decay at the colliders~\cite{Goldberger:1999wh,Davoudiasl:1999tf,Pomarol:1999ad,Chang:1999nh,Davoudiasl:2000wi}. Recently, a top-philic KK graviton model with non-universal couplings to SM fields~\cite{Geng:2018hpq} in which only the right-handed top-quarks interact strongly with the KK graviton has been considered as a possible solution to the gauge hierarchy problem. In this framework, the 5D warped space-time has two boundaries corresponding to UV and IR branes. The SM fields in this framework are 5D objects which either propagate in the bulk or are located on the branes. The interaction between the particles is given by the overlap of particle wave functions thereby making the interactions hierarchical. The first KK graviton is assumed to have a maximum near the IR brane so that it would couple strongly with the fields near the IR brane. It's coupling to SM particles would decrease exponentially depending on the variations of the SM wave function from the IR to the UV brane. Since the search for DM particles interacting non-gravitationally with SM particles has come under severe constraints from current observational data, one has been led to explore the interesting possibility of DM interacting with the SM particles by purely gravitational interaction in the RS framework~\cite{Kraml:2017atm,Han:2015cty,Lee:2013bua,Lee:2014caa,Rueter:2017nbk}.

It was observed that in most of the benchmark models considered in the literature, the desired values of the relic density could only be attained for the case of vector DM with mass near the $s$-channel KK graviton resonance~\cite{Rueter:2017nbk} or for very low Graviton and DM mass~\cite{Lee:2014caa} in a very narrow window in the parameter space. We consider here bulk RS model in which (i) only the right-handed top-quarks (top-philic) interact strongly with KK graviton~\cite{Lee:2014caa} or (ii) only the right-handed leptons (lepto-philic) interact strongly with the KK gravitons. We also extend the study to consider spin-3/2 DM particle, along with the scalar, vector and spin-1/2 fermionic DM particles.

In \cref{sec:model}, we describe our benchmark models in the RS framework. In \cref{sec:dmpheno}, we calculate the relic abundance through KK graviton exchange with DM particles. The constraints arising from the direct and indirect searches are obtained. \cref{sec:rc} is devoted to the results and conclusions. The decay width and DM annihilation cross-section expressions are given in the \cref{sec:a} and \cref{sec:b} respectively.

\section{Dark matter in the Randall-Sundrum framework}
\label{sec:model}
In the Randall-Sundrum framework, the particle interaction with the massive spin-2 gravitons $Y_{\mu\nu}$ is purely gravitational and is through the energy-momentum tensor being given by
\begin{align}
  {\cal L}_{\rm int} = - \sum \frac{c_i}{\Lambda} Y_{\mu\nu} T_i^{\mu\nu},
\end{align}
where $T_i^{\mu\nu}$ is the energy-momentum tensor of the $i^{\rm th}$ particle, $c_i$ is the corresponding coupling and $\Lambda$ is the scale of KK graviton interaction. We assume the DM fields to be either a real scalar, a real vector, a vector-like spin-1/2 Dirac fermion or a spin-3/2 fermion. The DM fields are further assumed to be SM singlets which do not carry any SM charge. It is also assumed that the DM particles are odd under a discrete ${\mathbb Z}_2$ symmetry so that there is no mixing between the DM and SM fields. It is assumed that the DM particles live on the IR brane. The mass of the DM particle is taken to be less than the mass of the gravitons for simplicity so that there is no DM annihilation into KK graviton states. We consider two benchmark models depending upon the relative placement of SM particles on or near the branes
\begin{description}
  \item[Model A] In the first benchmark model (Top-philic KK graviton) the right-handed top-quarks alone are assumed to be located on the IR brane, $SU(3)_C$ and $U(1)_Y$ gauge bosons live in the bulk and the rest of the SM fields including the $SU(2)_L$ gauge bosons live on the UV brane or close to it.
  \item[Model B] In the second benchmark model (lepto-philic KK graviton), the right-handed leptons are assumed to live on the IR brane and only the $U(1)_Y$ gauge bosons are assumed to live in the bulk with the rest of the SM particles including the gauge bosons live close to or on the UV brane.
\end{description}
In both the models discussed above the SM Higgs boson is placed far away from the IR brane and is thus incapable of solving  the hierarchy problem with the usual large warped factor in the original RS proposal. Also with the completely UV localised left-handed and IR localised right-handed top quarks, it may be difficult to generate the top quark mass. However, placement of left-handed and right-handed top quark fields strictly on the UV and IR branes is perhaps not required. The overlap of these two fields in the bulk can still generate top mass term, even though a large amount of tuning of 5D parameters is needed to achieve its observed large value~\cite{Pomarol:1999ad,Chang:1999nh}. Further, the spectrum of massive KK gluons and of $U(1)$ bulk gauge fields will have interactions with the SM fields on the IR brane. These KK gauge bosons are in general lighter than the KK graviton modes and interact strongly with the SM fields on the brane. The coupling can however, be somewhat suppressed with the use of brane localised kinetic terms. These comparatively strongly coupled gauge KK states will result in strong constraints on their masses from the electroweak precision data. This would result in large bulk curvature in severe tension with the solution of the hierarchy problem in the RS framework~\cite{Davoudiasl:2000wi,Davoudiasl:1999tf}.
However, our purpose here is to investigate the possibility of a DM particle interacting purely gravitationally in a non-universal RS frame work and still be able to generate the observed DM relic density without severally transgressing the constraints arising from the collider searches and from the direct and indirect experiments. We have also not considered radion exchange in our study since the radion coupling to matter is through the trace of the energy momentum tensor. The radion in general is lighter than the first KK graviton mass and its coupling to matter is reduced. The cross-section through radion exchange is thus suppressed away from the radion resonance. In our bench mark models however, the radion coupling to the right-handed fermions vanishes. The interaction Lagrangian relevant for the two benchmark models is thus given by: 
%
%In the present framework, the graviton alone is expected to interact strongly with particles on the IR brane while the coupling with $W^\pm$, SM Higgs boson and other fermion is weak. The coupling of graviton with the gauge bosons living in the bulk is suppressed by the volume factor of the order of $1/\ln(M_{Pl}/M_{IR}) \sim 0.03$ with the IR brane scale taken to be at ${\cal O}(\rm TeV)$ scale. This suppression factor is of the order of one-loop suppression factor $\alpha/4\pi$ for the EW or colour gauge bosons, $\alpha$ being the corresponding fine structure constant ($\alpha = \alpha_S$ for colour gluons and $\alpha_{\rm em}$ for electro-weak gauge bosons). The interaction Lagrangian relevant for the two benchmark models is this given by
%
\begin{description}[leftmargin=0pt,labelindent=0pt]
  \item[Model A]
  \begin{align}
    {\cal L}_{\rm int}
    =&\,
    - \frac{1}{\Lambda}
    \Big[ i \frac{c_{tt}}{4} \left\{ \bar t_R \left(\gamma^\mu \overleftrightarrow{D}^\nu + \gamma^\nu \overleftrightarrow{D}^\mu \right) t
    - 2\, \eta^{\mu\nu} \bar t_R \overleftrightarrow{\slashed D} t_R \right\}
    \nl
    &\,
    + \frac{\alpha_S}{4\pi} c_{gg}
    \Big\{\frac{1}{4} \eta^{\mu\nu} G_a^{\lambda \rho} G^a_{\lambda \rho} - G_a^{\mu\lambda} G^{a\nu}_{\lambda} \Big\}
    + \frac{\alpha}{4\pi} c_1 \Big\{ \frac{1}{4}\eta^{\mu\nu} B^{\lambda\rho} B^{\lambda\rho} - B^{\mu\lambda} B^{\nu}_{\lambda} \Big\}
    \Big]\,Y_{\mu\nu}, \label{lintA}
  \end{align}
  where $D_\mu = \d_\mu + i (2/3)\,g_1 B_\mu + i g_s t^a G^a_\mu$ is the covariant derivative for the right-handed top quark field, $G_\mu$ and $B_\mu^a$ are the gluon and $U(1)_Y$ gauge boson fields respectively. The coupling $c_{tt}$ are scaled by appropriate loop suppression and $c_{tt} > c_{gg} \alpha_S/4\pi \sim c_1 \alpha/4\pi \gg$ other couplings.

  \item[Model B]
  \begin{align}
    {\cal L}_{\rm int}
    =&\,
    - \frac{1}{\Lambda}
    \Big[ i \sum_{\ell=e,\mu,\tau} \frac{c_{\ell\ell}}{4} \left\{ \bar\ell_R \left(\gamma^\mu \overleftrightarrow{D}^\nu + \gamma^\nu \overleftrightarrow{D}^\mu \right) \ell
    - 2\, \eta^{\mu\nu} \bar\ell_R \overleftrightarrow{\slashed D} \ell_R \right\}
    \nl
    &\,
    + \frac{\alpha}{4\pi} c_1 \Big\{ \frac{1}{4}\eta^{\mu\nu} B^{\lambda\rho} B^{\lambda\rho} - B^{\mu\lambda} B^{\nu}_{\lambda} \Big\}
    \Big]\,Y_{\mu\nu}, \label{lintB}
  \end{align}
  where $D_\mu = \d_\mu + i \frac{2}{3} g_1 B_\mu$ is the covariant derivative for the right-handed leptons.
\end{description}
The DM particles scalars ($S$), vectors ($V$), spin-1/2 fermions ($\chi$) and spin-3/2 fermions ($\Psi$) are taken to be present on the IR brane and interact with KK gravitons with a coupling strength of order one, through the energy-momentum tensor. The interaction Lagrangian is given as
\begin{align}
  {\cal L}_{\rm int} = -\frac{1}{\Lambda} c_{\rm DM} Y_{\mu\nu} T_{\rm DM}^{\mu\nu},
\end{align}
where
\begin{align}
  T^{\mu\nu}_S
  =&\,
  (\d^\mu S)(\d_\mu S) - \frac{1}{2} \eta^{\mu\nu} \left[ (\d^\alpha S)(\d_\alpha S) - m_S^2 S^2 \right],
  \\
  T^{\mu\nu}_\chi
  =&\,
  \frac{i}{4} \bar\chi \left[ \gamma^\mu \overleftrightarrow{\d}^\nu + \gamma^\nu \overleftrightarrow{\d}^\mu \right] \chi
  - \eta^{\mu\nu} \left[ i \bar\chi {\slashed \gamma} \chi - m_{\chi} \bar\chi \chi \right],
  \\
  T^{\mu\nu}_V
  =&\,
  - V^{\mu\alpha}V_{\alpha}^{\nu} + m_V^2 V^\mu V^\nu
  + \eta^{\mu\nu} \left[ \frac{1}{4} V_{\alpha\beta}V^{\alpha\beta} - \frac{1}{2} m_V^2 V_\alpha V^\alpha \right],
  \\
  T^{\mu\nu}_\Psi
  =&\,
  \frac{i}{4}\,\bar\Psi_\alpha \left[ \gamma^\mu \overleftrightarrow{\d}^\nu + \gamma^\nu \overleftrightarrow{\d}^\mu \right] \Psi^\alpha
  - \frac{i}{4}\,\bar\Psi_\alpha \left[ \gamma^\mu \overleftrightarrow{\d}^\alpha \Psi^\nu + \gamma^\nu \overleftrightarrow{\d}^\alpha \Psi^\mu \right]\,.
\end{align}
In above energy-momentum tensor, $V_{\alpha\beta}$ is the vector field tensor. The spin-3/2 field $\Psi_\alpha$ satisfies the Euler-Lagrange equation
\begin{align}
  \left( i \slashed\d - m_\Psi \right) \Psi_\mu = 0,
\end{align}
with $\d^\nu \Psi_\nu = 0$ and $\gamma^\nu \Psi_\nu = 0$.

The spin-2 KK graviton polarisation sum is given by
\begin{gather}
  \sum_{s=1}^{5} \epsilon_{\mu\nu} (s)\epsilon_{\alpha\beta} (s)
  =\,
  \Pi^2_{\mu\nu,\alpha\beta} (p)
  =
  \frac{1}{2} \Pi^1_{\mu\alpha}\Pi^1_{\nu\beta}
  + \frac{1}{2} \Pi^1_{\mu\beta}\Pi^1_{\nu\alpha}
  - \frac{1}{3} \Pi^1_{\mu\nu}\Pi^1_{\alpha\beta} .
  \\
  \Pi^1_{\mu\nu}(p)
  =\,
  - \eta_{\mu\nu} + \frac{p_\mu p_\nu}{m^2_{Y}} ,
\end{gather}
$p^\mu \Pi^2_{\mu\nu,\alpha\beta} = 0$ and $\Pi^2_{\mu\nu,\alpha\beta} \eta^{\mu\nu} = 0$ for on-shell $Y^{\mu\nu}$.

The polarisation sum of spin-3/2 fermions
\begin{align}
  \Pi^{\mu\nu}
  =&\,
  \sum_{s=-3/2}^{\sfrac{3}{2}} u^\mu_{3/2}(s){\bar u}^\nu_{3/2}(s)
  \nl
  =&\,
  - \left( {\slashed p} + m_\Psi\right)
  \left[ \eta^{\mu\nu} - \frac{2}{3} \frac{p^\mu p^\nu}{m_\Psi^2} - \frac{1}{3}\gamma^\mu \gamma^\nu - \frac{1}{3} \frac{\left(p^\nu \gamma^\mu - p^\mu \gamma^\nu\right)}{m_{\Psi}}\right],
  \\
  \bar\Pi^{\mu\nu}
  =&\,
  \sum_{s=-3/2}^{\sfrac{3}{2}} v^\mu_{3/2}(s){\bar v}^\nu_{3/2}(s)
  = \Pi^{\mu\nu}\left(m_\Psi \to - m_\Psi \right).
\end{align}

After electroweak symmetry breaking, the coupling between $Y_{\mu\nu}$ and $U(1)_Y$ gauge bosons is written in terms of coupling of $Y_{\mu\nu}$ with photons and $Z$ bosons as
\begin{gather}
  {\cal L}_{Y}
  \supset\,
 - \frac{1}{\Lambda} \Bigg[ \frac{\alpha}{4\pi} c_{\gamma\gamma} \left(\frac{1}{4}\eta^{\mu\nu} A^{\lambda\rho}A_{\lambda\rho} -  A^{\mu\lambda}A_{\lambda}^\nu \right) + \frac{\alpha}{4\pi} c_{Z\gamma} \left(\frac{1}{4}\eta^{\mu\nu} A^{\lambda\rho}Z_{\lambda\rho} -  A^{\mu\lambda}Z_{\lambda}^\nu \right)
  \nl
  + \frac{\alpha}{4\pi} c_{ZZ} \left(\frac{1}{4}\eta^{\mu\nu} Z^{\lambda\rho}Z_{\lambda\rho} -  Z^{\mu\lambda}Z_{\lambda}^\nu \right) \Bigg] Y_{\mu\nu}.
\end{gather}
The couplings of $c_{\gamma\gamma}, c_{\gamma Z}$ and $c_{ZZ}$ can be obtained from the coupling $c_1$ and are given as
\begin{gather}
\begin{rcases}
  c_{\gamma\gamma} = c_1 \cos^2\theta_W ,
  \\
  c_{Z\gamma} = - c_1 \sin 2\theta_W = - \sin 2\theta_W c_{\gamma\gamma}/ \cos^2\theta_W ,
  \\
  c_{ZZ} = c_1 \sin^2\theta_W = \tan^2\theta_W c_{\gamma\gamma} .
\end{rcases}
\end{gather}

Since the gravitons couple strongly with the right-handed top quarks in benchmark model {\bf A} and with right-handed charged leptons in model {\bf B}, the top quark and lepton triangle loop contribution to the graviton-gluon and graviton-$U(1)_Y$ gauge bosons can typically be of the same order as the corresponding tree-level couplings. The resulting effective couplings are evaluated in Refs.~\cite{Geng:2016xin,Geng:2018hpq} and are given \cref{sec:a} where KK graviton decay expressions are also listed.

\begin{figure*}[!ht]
  \centering
  \subfloat[]{\includegraphics[width=0.48\linewidth]{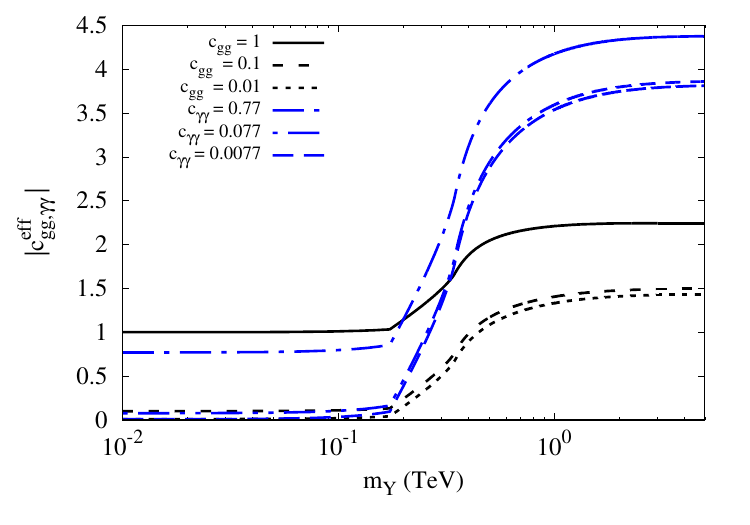}\label{fig:ceffA}}
  \hfill
  \subfloat[]{\includegraphics[width=0.48\linewidth]{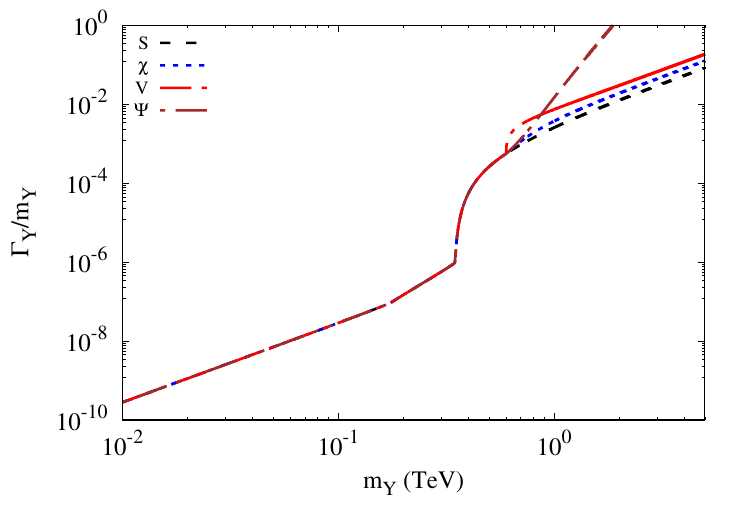}\label{fig:decayAb}}
  \\
  \subfloat[]{\includegraphics[width=0.48\linewidth]{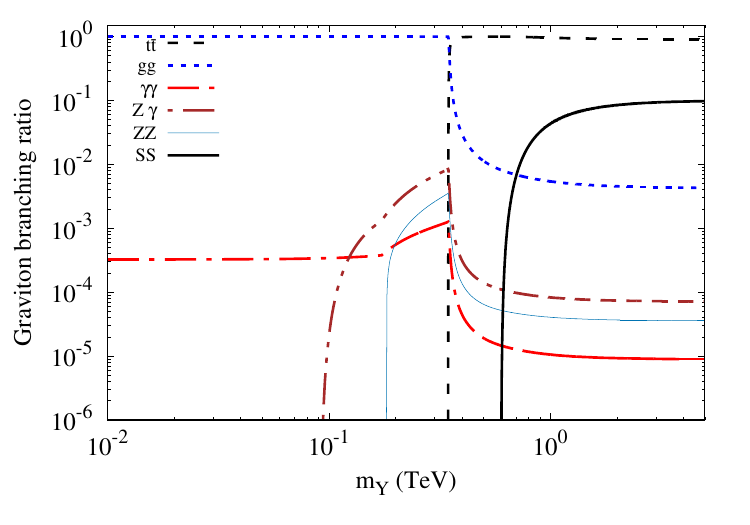}\label{fig:brscalarA}}
  \hfill
  \subfloat[]{\includegraphics[width=0.48\linewidth]{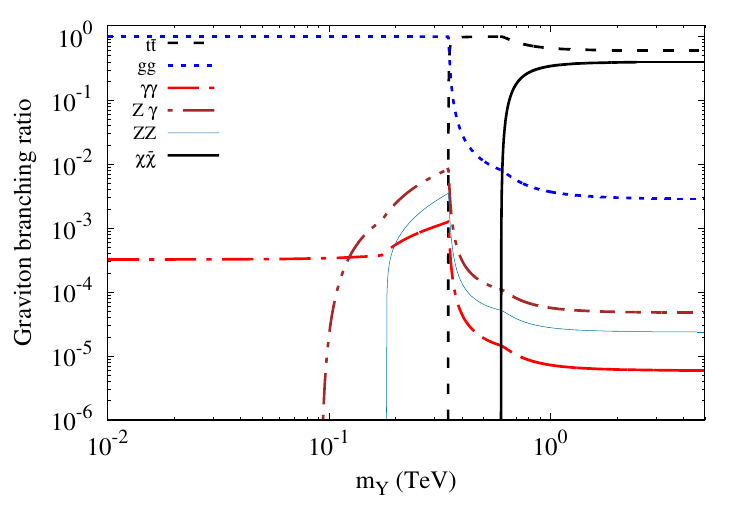}\label{fig:brspinorA}}
  \\
  \subfloat[]{\includegraphics[width=0.48\linewidth]{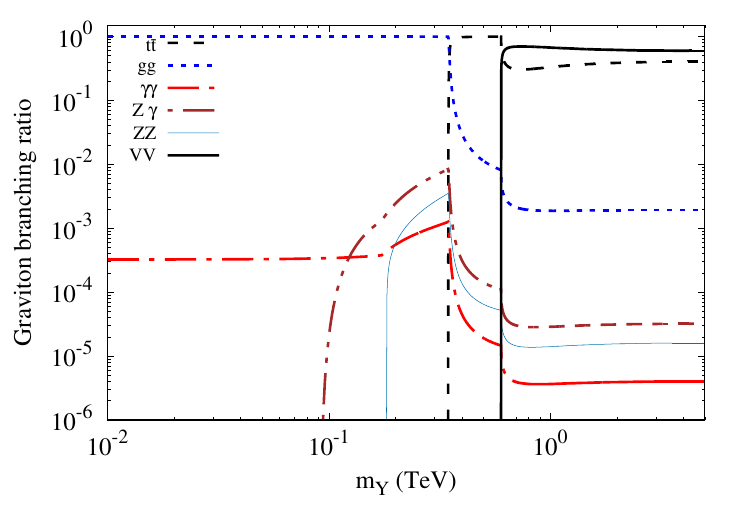}\label{fig:brvectorA}}
  \hfill
  \subfloat[]{\includegraphics[width=0.48\linewidth]{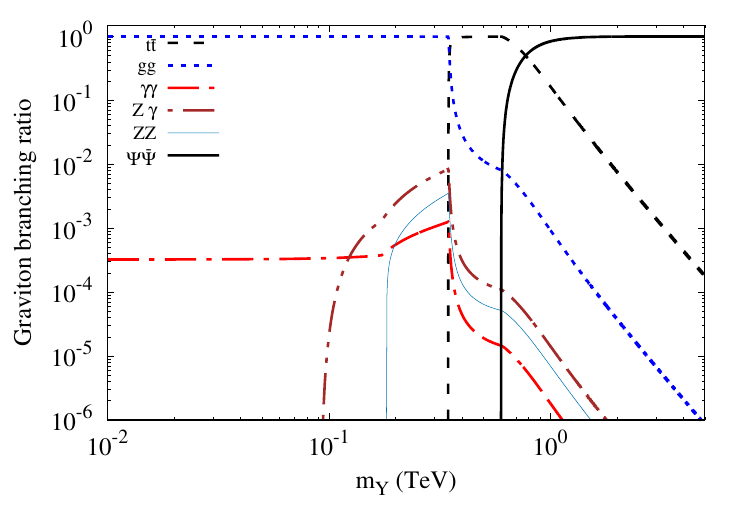}\label{fig:brraritaA}}
  \caption{KK-graviton-gauge boson effective couplings, decay width branching ratios in the final states relevant to the bench-mark model {\bf A} . The couplings $c_{gg}$, $c_{tt}$ and $c_1$ defined in \cref{lintA} are taken to be equal to~1 ($\left| c^{\rm eff}_{gg,\gamma\gamma}\right|$ is shown for different values too). The KK-graviton interaction scale factor $\Lambda$ is fixed at 1~TeV and the dark matter mass at 300~GeV. Panels~(\ref{fig:brscalarA}~-~\ref{fig:brraritaA}) show the branching ratios separately for the scalar, spin-1/2, vector and spin-3/2 dark matter.}
  \label{fig:decayA}
\end{figure*}
\begin{figure*}[!ht]
  \centering
  \subfloat[]{\includegraphics[width=0.48\linewidth]{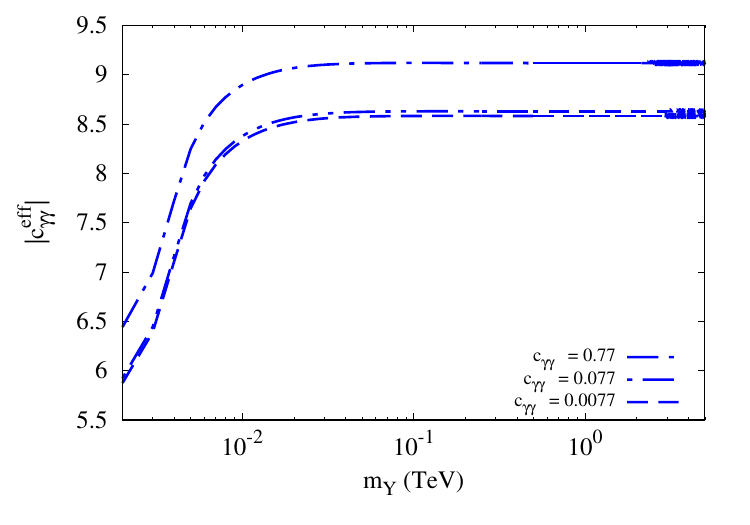}\label{fig:ceffB}}
  \hfill
  \subfloat[]{\includegraphics[width=0.48\linewidth]{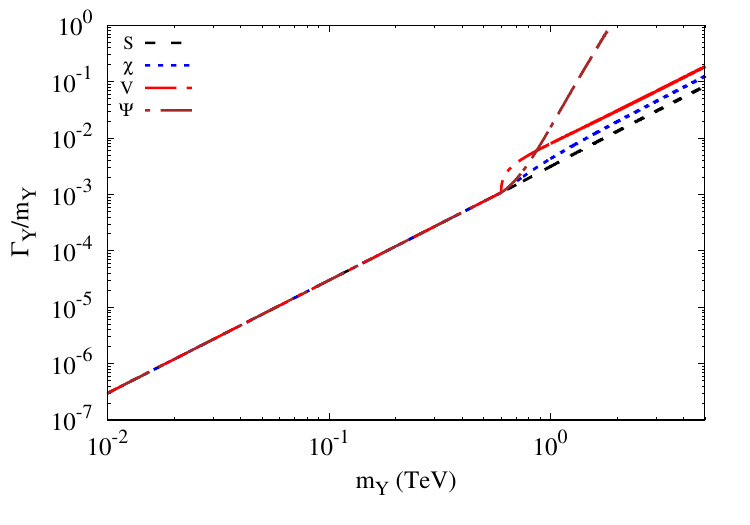}\label{fig:decayBb}}
  \\
  \subfloat[]{\includegraphics[width=0.48\linewidth]{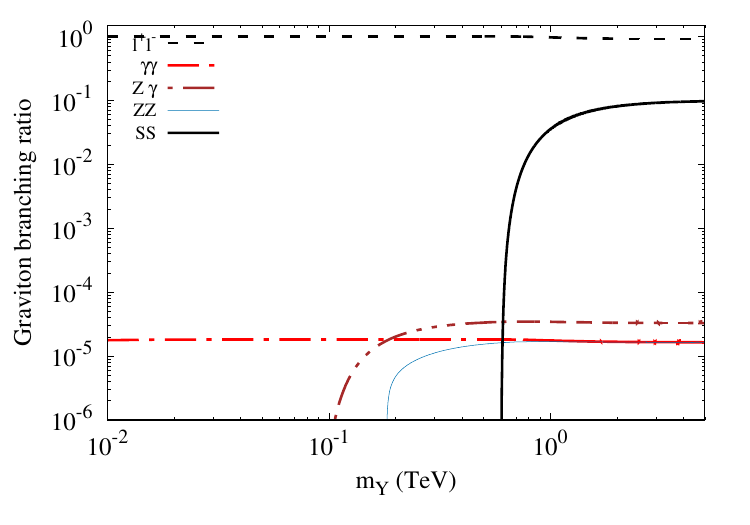}\label{fig:brscalarB}}
  \hfill
  \subfloat[]{\includegraphics[width=0.48\linewidth]{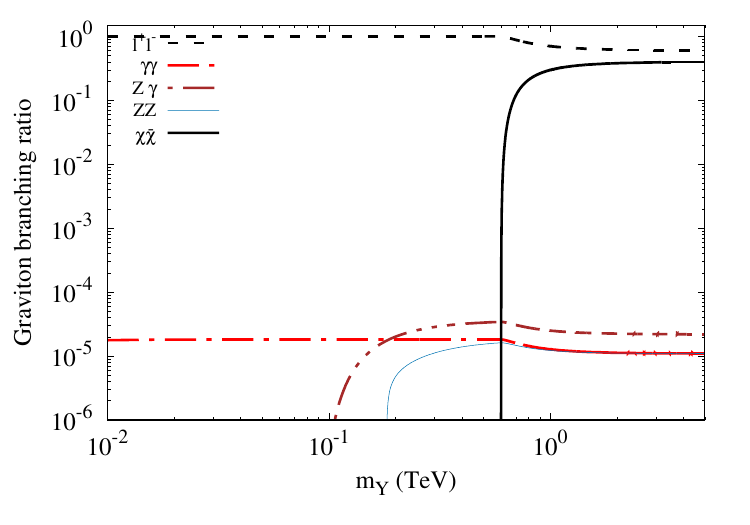}\label{fig:brspinorB}}
  \\
  \subfloat[]{\includegraphics[width=0.48\linewidth]{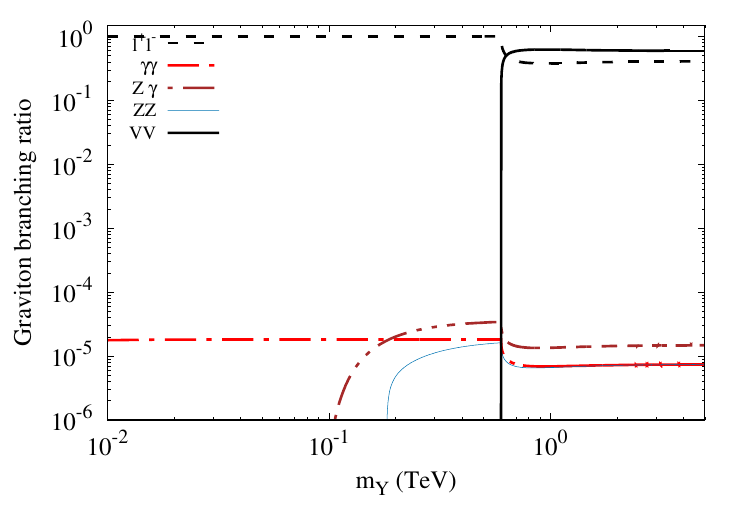}\label{fig:brvectorB}}
  \hfill
  \subfloat[]{\includegraphics[width=0.48\linewidth]{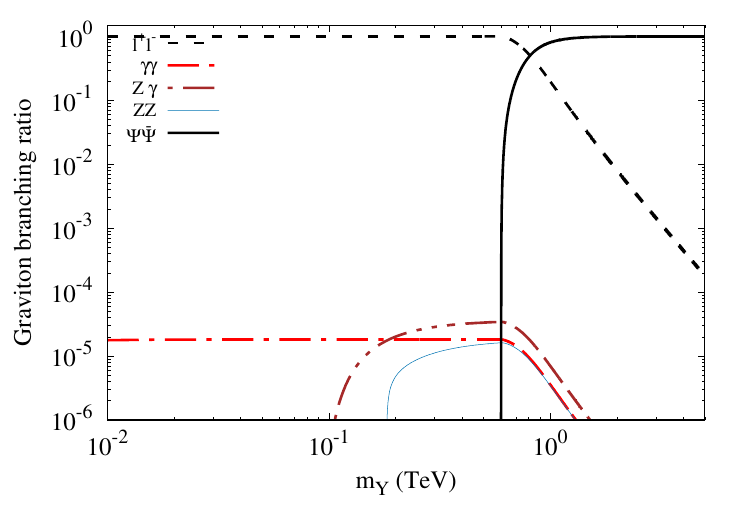}\label{fig:brraritaB}}
  \caption{KK-graviton-gauge boson effective couplings, decay width   branching ratios in the final states relevant to the bench-mark model {\bf B}. The couplings $c_{ll}$ and $c_1$ defined in \cref{lintB} are taken to be equal to~1 ($\left| c^{\rm eff}_{\gamma\gamma}\right|$ is shown for different values too). The KK-graviton interaction scale factor $\Lambda$ is fixed at 1~TeV and the dark matter mass at 300~GeV. Panels~(\ref{fig:brscalarB}~-~\ref{fig:brraritaB}) show the branching ratios separately for the scalar, spin-1/2, vector and spin-3/2 dark matter.}
  \label{fig:decayB}
\end{figure*}
It may be mentioned that $Y \to W^+W^-$ is not considered here. This is because  the coupling $YW^+W^-$ is induced at the one loop level through the right-handed top quark and the right-handed leptonic loop in models {\bf A} and {\bf B} respectively. The coupling  is further suppressed in comparison to even the $Y \gamma \gamma$ coupling by an extra factor of $(m_t/m_Y)^2$~\cite{Geng:2016xin} in model {\bf A} and by a factor $(m_l/m_Y)^2$ which is vanishingly small in the latter case. In model {\bf A} the suppression may be comparable to $Y \gamma \gamma$, $Y Z \gamma$ or $Y Z Z$ coupling for at least low graviton mass which in our analysis lies between 200~GeV to several~TeV. The branching ratio of $Y \to \gamma \gamma$ is already several orders of magnitude less than into the dominant right-handed top quark and right-handed lepton pairs in models {\bf A} and {\bf B} respectively as can be decerned from~\cref{fig:decayA,fig:decayB}.  

In~\cref{fig:decayA,fig:decayB} we have plotted the KK-graviton-gauge boson effective couplings, decay width and branching ratios in the final states relevant to the benchmark models {\bf A} and {\bf B} respectively. For the purpose of illustration we have taken the couplings $c_{gg}$, $c_{tt}$, $c_{ll}$ and $c_1$ defined in \cref{lintA,lintB} to be equal to~1. The KK-graviton interaction scale factor $\Lambda$ is fixed at 1~TeV and the dark matter mass is taken to be equal to 300~GeV. We see from~\cref{fig:decayA,fig:decayB} that after the onset of KK-graviton decay into DM particles at 600~GeV, the decay width for the case of spin-3/2 DM particle rises rapidly with the increase in graviton mass $m_Y$ in comparison to the case of scalar, spin-1/2 and vector DM particles. This is a general feature of spin-3/2 particles~\cite{Khojali:2016pvu, Khojali:2017tuv} and can be seen from the decay width expression~(\ref{decay-kk}) of  $Y$ into $\Psi \bar \Psi$. In~\cref{fig:brscalarA,fig:brspinorA,fig:brvectorA,fig:brraritaA} we have shown the branching ratios separately for the scalar, spin-1/2, vector and spin-3/2 dark matter in the benchmark model {\bf A} and likewise for the model {\bf B} in~\cref{fig:brscalarB,fig:brspinorB,fig:brvectorB,fig:brraritaB}. We note that the graviton branching ratio into spin-3/2 DM pairs approaches one with the increase in graviton mass.

\begin{figure*}[!ht]
  \centering
  \subfloat[]{\includegraphics[width=0.48\linewidth]{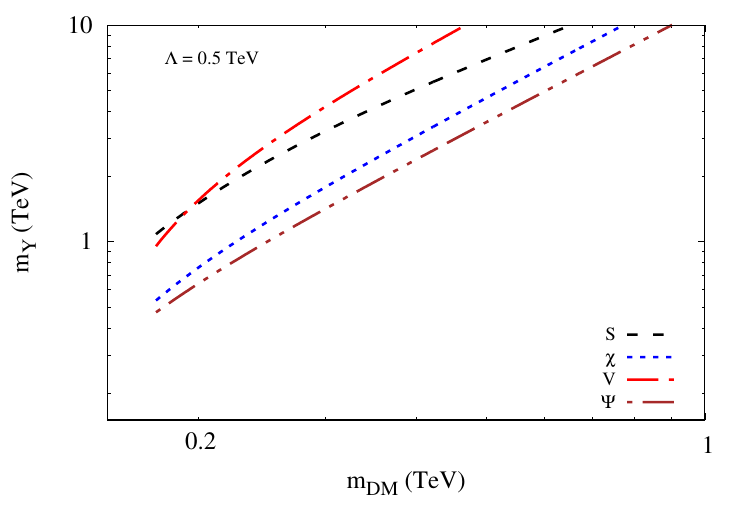}\label{fig:relicA500}}
  \hfill
  \subfloat[]{\includegraphics[width=0.48\linewidth]{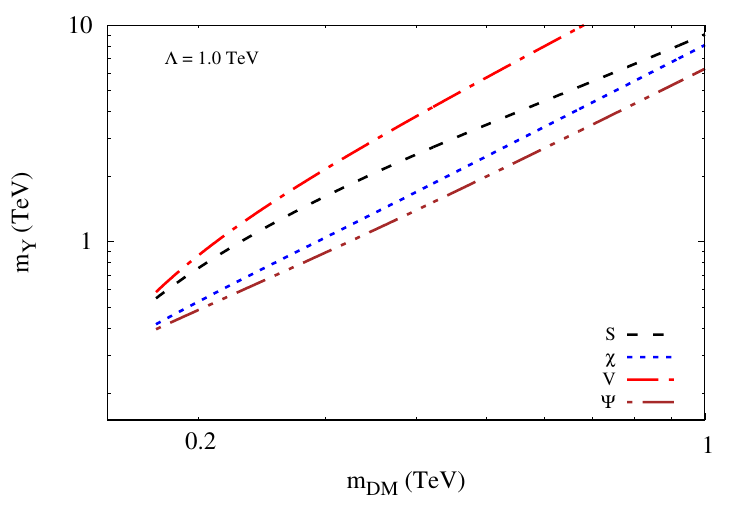}\label{fig:relicA1000}}
  \\
  \subfloat[]{\includegraphics[width=0.48\linewidth]{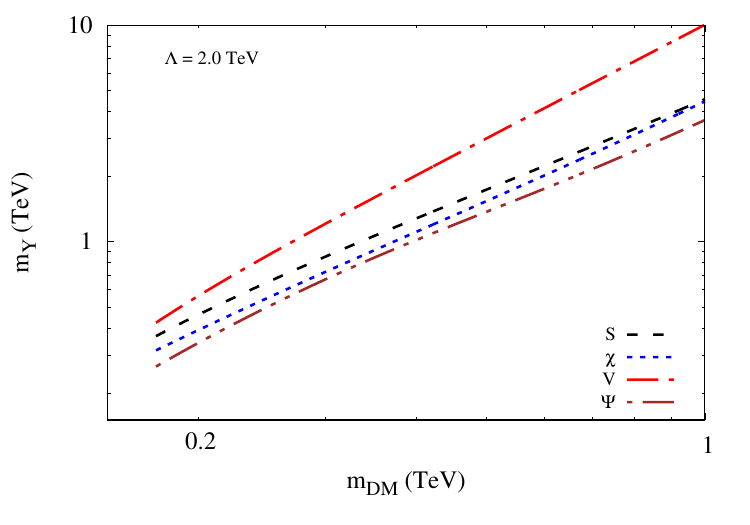}\label{fig:relicA2000}}
  \hfill
  \subfloat[]{\includegraphics[width=0.48\linewidth]{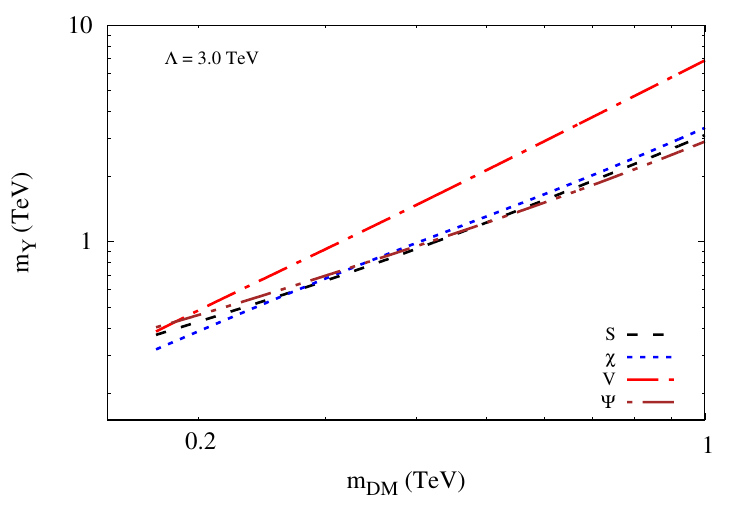}\label{fig:relicA3000}}
  \caption{Contours of constant relic density $\Omega_{\rm DM} h^2 = 0.119$ in the $m_{\rm DM} - m_Y$ plane for the case of scalar, spin-1/2, vector and spin-3/2 dark matter particles in the benchmark model {\bf A}. The panels~(\ref{fig:relicA500}~-~\ref{fig:relicA3000}) correspond to the KK-graviton interaction scale $\Lambda = 500$~GeV, 1~TeV, 2~TeV and 3~TeV respectively. The value of couplings $c_{tt}$, $c_{gg}$ and $c_1$ are taken to be equal to~1.}
  \label{fig:relicA}
\end{figure*}

\begin{figure*}[!ht]
  \centering
  \subfloat[]{\includegraphics[width=0.48\linewidth]{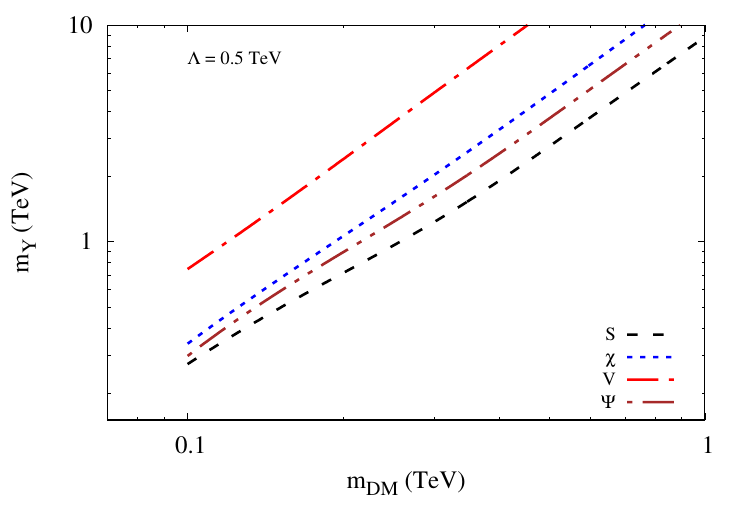}\label{fig:relicB500}}
  \hfill
  \subfloat[]{\includegraphics[width=0.48\linewidth]{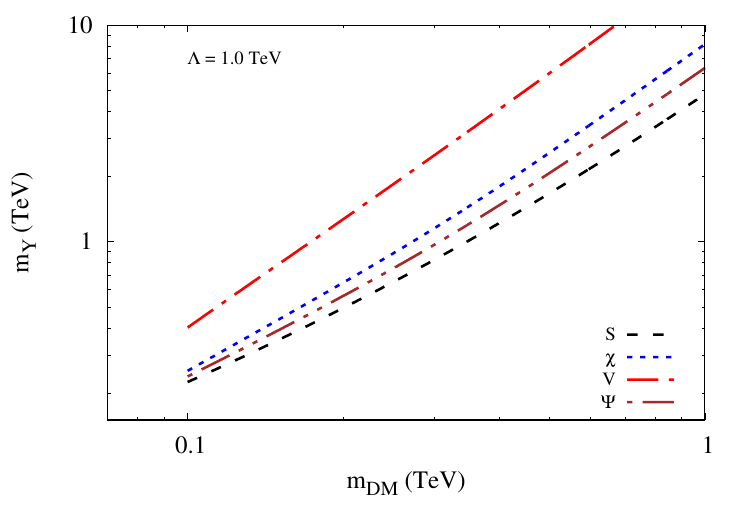}\label{fig:relicB1000}}
  \\
  \subfloat[]{\includegraphics[width=0.48\linewidth]{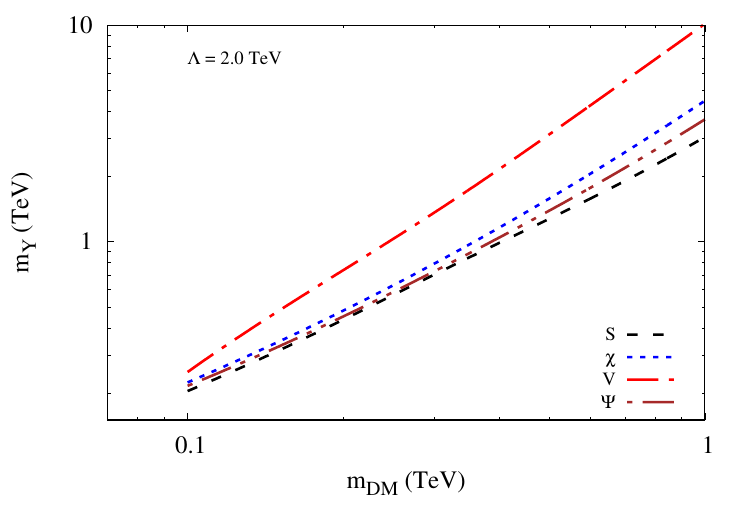}\label{fig:relicB2000}}
  \hfill
  \subfloat[]{\includegraphics[width=0.48\linewidth]{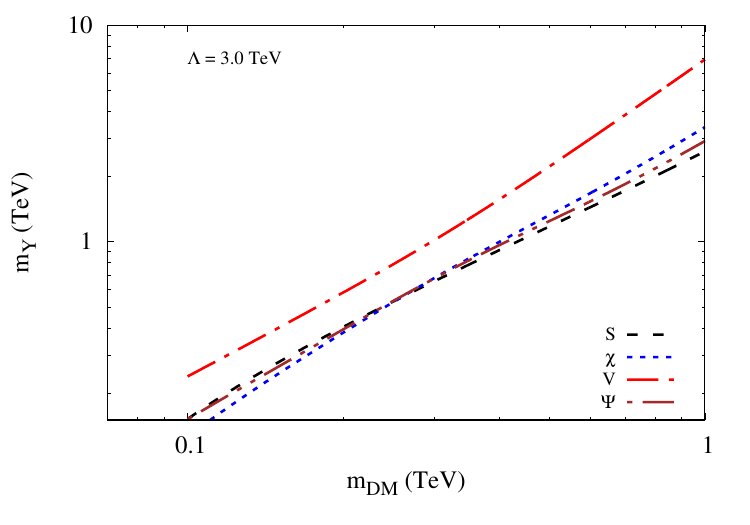}\label{fig:relicB3000}}
  \caption{Contours of constant relic density $\Omega_{\rm DM} h^2 = 0.119$ in the $m_{\rm DM} - m_Y$ plane for the case of scalar, spin-1/2, vector and spin-3/2 dark matter particles in the benchmark model {\bf B}. The panels~(\ref{fig:relicB500}~-~\ref{fig:relicB3000}) correspond to the KK-graviton interaction scale $\Lambda = 500$~GeV, 1~TeV, 2~TeV and 3~TeV respectively. The value of couplings $c_{ll}$ and $c_1$ are taken to be equal to~1.}
  \label{fig:relicB}
\end{figure*}
\section{Dark matter phenomenology}
\label{sec:dmpheno}
\subsection{Thermal relic density}
\label{sec:relic}
The Randall-Sundram model with its warped compactification of extra space dimensions can have profound cosmological ramifications involving phase transition to an inflationary phase at low temperatures of the order of weak scale and on the thermal history of WIMPS. The expansion of the Universe in the RS  model in the presence of radius stabilisation mechanism however, turns out to be in agreement with the effective 4-D description~\cite{Creminelli:2001th, Csaki:1999mp}. Thus in the early Universe, DM is in thermal equilibrium with the hot dense plasma and as the Universe expands and cools, it freezes out. The evolution of the number density of DM matter $n_{\rm DM}$ is governed by the Boltzmann equation
\begin{align}
  \frac{d n_{\rm DM}\nquad}{dt}\quad
  =
  - 3\,\frac{{\dot a}}{a}\,n_{\rm DM} - \braket{\sigma | v} \big[ n_{\rm DM}^2 - \left( n_{\rm DM}^{\rm eq}\right)^2 \big] \,.
\end{align}
$n_{\rm DM}^{\rm eq}$ is the equilibrium density of the DM at temperature $T$ and is given by
\begin{align}
  n_{\rm DM}^{\rm eq} = g \left( \frac{m_{\rm DM} T}{2\pi}\right)^{\sfrac{3}{2}} \exp\left[-\frac{m_{\rm DM}\nquad}{T}\quad\right] \,,
\end{align}
where $g$ is the number of degrees of freedom ($g = 1, 2, 3$) and $4$ respectively for scalar, spin-1/2, vector and spin-3/2 fermionic DM. $\braket{\sigma | v}$ is the thermal averaged cross-section of DM annihilation channels into the SM states. The thermally averaged cross-section can be written as a one-dimensional integral over the centre of mass energy square $s$ as
\begin{align}
  \braket{\sigma | v}
  =
  \frac{1}{8\,m_{\rm DM}^2 \left[{\cal K}_2\left( m_{\rm DM}/T\right)\right]^2} \int\limits_{4 m_{\rm DM}^2}^\infty \left(s - 4 m_{\rm DM}^2 \right) \sqrt{s} \,\sigma_{\rm ann}\,{\cal K}_1 \left(\frac{\sqrt{s}}{T} \right) ds \,,
\end{align}
where ${\cal K}_1$ and ${\cal K}_2$ are the modified Bessel functions of the second kind, the annihilation cross-section $\sigma_{\rm ann}$ depends only on the masses, $s$ and couplings of the DM and SM particles involved.

The Boltzmann equation can be solved to give the thermal relic density
\begin{align}
 \Omega_{\rm DM} h^2
 \simeq
 \frac{1.07 \times 10^9\, x_F}{M_{\rm Pl} \sqrt{g^*(x_F)} \,\braket{\sigma | v}} \,.
\end{align}
Here $h$ is the Hubble parameter today, $g^*(x_F)$ is total number of dynamic degrees of freedom near the freeze-out temperature $T_F$ and $x_F = m_{\rm DM} / T_F$ is obtained by solving
\begin{align} 
  x_F
  =
  \ln \left[ c(c+2) \sqrt{\frac{15}{8}} \frac{g\,M_{\rm Pl}\,m_{\rm DM} \braket{\sigma | v}}{8 \pi^3\,\sqrt{x_F}\,\sqrt{g^*(x_F)} } \right]
\end{align}
and $c$ is of order~1. Scattering cross-sections as a function of $s$ and DM mass $m_{\rm DM}$ and couplings are given in the \cref{sec:b}. Since the mediator KK graviton state carries spin-2, the annihilation rates into final SM states in the benchmark models considered here, are velocity suppressed. The suppression depends on the spin of the initial DM and final SM particles. The maximum suppression occurs for the scalar and spin-1/2 DM's and there is no suppression for the vector and spin-3/2 DM particles in the dominant annihilation channels. We first examine the benchmark model {\bf A} (top-philic) in which only the right-handed top quarks and the DM particles live on the IR-brane.

We have computed the relic density in the two benchmark models considered here numerically. We have generated the input model files using \textsc{LanHEP}~\cite{Semenov:2014rea} which calculates all the required couplings and Feynman rules by using the Lagrangian given in \cref{sec:model}. Then the model files are used in \textsc{CalcHEP}~\cite{Belyaev:2012qa} to calculate the required decay widths annihilation cross sections. In~\cref{fig:relicA,fig:relicB} we show the $2\sigma$ contour of constant relic density $0.119$ in the DM mass ($m_{\rm DM}$) and KK graviton mass ($m_Y$) for fixed KK graviton interaction scale $\Lambda$.

In~\cref{fig:relicA,fig:relicB}, we have shown the contours of constant relic density $\Omega_{\rm DM} h^2 = 0.119$ in the $m_{\rm DM} - m_Y$ plane for the case of scalar, spin-1/2, vector and spin-3/2 dark matter particles in the benchmark models {\bf A} and {\bf B} respectively. 
The couplings $c_{tt}, c_{gg}, c_{\gamma \gamma}$ and  $c_{ll}$ are chosen to be equal to $1$ and four different values of the KK graviton interaction scale $\Lambda$ are taken to be 500~GeV, 1~TeV, 2~TeV and 3~TeV respectively. The dark matter annihilation cross-sections scale inversely with $\Lambda^4$ and directly with the square of $c_{tt}, c_{ll}, c_{gg}^{\rm eff}$ and $c_{\gamma \gamma}^{\rm eff}$ couplings. The  $c_{gg}^{\rm eff}$ and $c_{\gamma \gamma}^{\rm eff}$ couplings are expressed in terms of bare $c_{gg}$ and $c_{\gamma \gamma}$ couplings in~\cref{eq:ceffGGA}-(\ref{eq:ceffggA}) and~\cref{eq:ceffggB} for benchmark models {\bf A} and {\bf B} respectively. We notice that the panels~(\ref{fig:relicA500}~-~\ref{fig:relicA3000}) and (\ref{fig:relicB500}~-~\ref{fig:relicB3000}) correspond to the KK-graviton interaction scale  $\Lambda = 500$~GeV, 1~TeV, 2~TeV and 3~TeV respectively. It can be seen from these figures that the distinction with respect to the contribution to the relic density arising from the scalar, spin-1/2, vector or spin-3/2 dark matter particles tends to diminish with the increase in the interaction scale $\Lambda$. We have also found that a decrease in the couplings $c_{gg}$ and $c_{\gamma \gamma}$ by as much as  one or even two orders of magnitude does not have a noticeable effect on the relic density curves. This can be easily understood from the effective coupling expressions in the~\cref{sec:a} and the fact that the dominant DM annihilation processes in the models {\bf A} and {\bf B} are to the right-handed top quark and lepton pairs respectively.

\subsection{Direct detection}
\label{sec:dd}
The tree-level nucleon-DM scattering through the massive spin-2 KK graviton propagator, after integrating out the massive graviton field tends to the effective Lagrangian
\begin{align}
  {\cal L}_{\rm eff}
  =
  i \frac{c_{\rm DM}\,c_{\rm SM}}{2\,\Lambda^2\,m_Y^2} \left[2 {\tilde T}_{\mu\nu}^{\rm DM}\,{\tilde T}^{\mu\nu}_{\rm SM} - \frac{1}{6} {T}^{\rm DM}{T}_{\rm SM} \right],
\end{align}
where ${\tilde T}_{\mu\nu}$ and $T$ are traceless and trace part of the energy-momentum tensor {\it viz.}
\begin{align}
  T_{\mu\nu}
  =
  {\tilde T}_{\mu\nu} + \frac{1}{4}\eta_{\mu\nu} T; \qquad T = \eta^{\mu\nu} T_{\mu\nu}.
\end{align}
For the DM-nucleon scattering, the relevant SM energy-momentum tensor is the one that involves light quarks and gluons. In the top-philic RS model {\bf A} considered here,
\begin{align}
  T^{\rm SM}_{\mu\nu}
  =
  T^g_{\mu\nu} = \frac{\alpha_S}{4 \pi} \left[ \frac{1}{4} \eta_{\mu\nu} G_a^{\lambda \rho} G^a_{\lambda \rho} - G^a_{\mu\lambda } G_{a\nu}^{\lambda}  \right] \,.
\end{align}
The trace and traceless part of $T^a_{\mu\nu}$ are given by
\begin{align}
  T^g
  =
  \frac{\alpha_S}{4 \pi} G^a_{\mu\nu} G^{\mu\nu}_a \quad {\rm and} \quad
  T^g_{\mu\nu}
  =
  \frac{\alpha_S}{4 \pi} \left[ G^{a\rho}_{\mu} G_{a\rho\nu} - \frac{1}{4}\eta_{\mu\nu} G^a_{\rho\sigma} G_a^{\rho\sigma} \right]
\end{align}
respectively. The matrix element of the trace part of the energy-momentum tensor $T^g_{\mu\nu}$ between nucleon and state can be evaluated easily~\cite{Alarcon:2011zs,Hisano:2015bma} and are given by
\begin{align}
  \bra{N} T^g \ket{N}
  =
  \bra{N} \frac{\alpha_S}{4\pi} G^a_{\mu\nu} G_a^{\mu\nu} \ket{N}
  =
  - \frac{8}{9\times4}\,m_N\,f_{TY}^N,
\end{align}
where
\[
  f_{TY}^{(N)} = 1 - \sum_{q=u,d,s} f_{Tq}^N \simeq 0.92 \,.
\]
The nucleon-matrix element of the traceless part of the energy-momentum tensor is given by the second moment of the gluon distribution function of the factorisation scale.
\begin{align}
  &\bra{N} \frac{\alpha_S}{4 \pi} \left[ G^{a\rho}_{\mu} G_{a\rho\nu} - \frac{1}{4}\eta_{\mu\nu} G^a_{\rho\sigma} G_a^{\rho\sigma} \right] \ket{N}
\nl
  &\qquad\qquad\qquad=
  - \frac{\alpha_S}{4 \pi}\frac{1}{m_N} \left[ p_\mu p_\nu - \frac{1}{4} m_N^2 \eta_{\mu\nu} \right] \left[f_{TY}^{(N)}\right]^2 g\left(2:\mu\right),
\end{align}
where $g\left(2:\mu\right) = \int\limits_0^1 x\, g_2\left(x,\mu \right) dx$ and is $\sim 0.464$ and the matrix element is suppressed by $\alpha_S/{4\pi}$ in comparison to the matrix element of the trace part of the energy-momentum tensor between nucleon states. The contribution of the traceless part of the energy-momentum tensor to the nucleon-DM scattering turns out to be $\sim 10$\% compared to the trace contribution. Thus in the N-R limit, neglecting small momentum dependent terms, the DM-nucleon scattering in the lab frame can be calculated from the trace part of the energy-momentum tensor. In this approximation, the spin-independent scattering cross-section does not depend on the spin of the DM particle. The spin-independent DM-nucleon scattering cross-section is given by
\begin{align}
  \sigma_{\rm SI}
  =&\,
  \frac{1}{729 \pi} \left(c_{gg}^{\rm eff}\right)^2 c_{\rm DM}^2 \left(\frac{\mu}{m_N}\right)^2 \left(\frac{m_N}{\Lambda\,m_Y} \right)^4 m^2_{\rm DM} \left[f_{TY}^{(N)}\right]^2
  \nl
  \simeq
  &\,
  1.75\times 10^{-49} \left(c_{gg}^{\rm eff} \right)^2 c_{\rm DM}^2  \left(\frac{m_{\rm DM}}{\rm TeV}\right)^2 \left(\frac{\rm TeV}{\Lambda}  \right)^4 \left(\frac{\rm TeV}{m_Y}  \right)^4\,{\rm cm^2}, {\label{sigmasi}}
\end{align}
where $m_N$ the nucleon mass and $\mu = \frac{m_{\rm DM} m_N}{m_{\rm DM} + m_N}$ is the reduced mass.

In the lepto-philic model {\bf B} considered here, DM-nucleon scattering arises through KK graviton-photon effective coupling which is suppressed by $\alpha/{4\pi}$. The DM-nucleon scattering cross-section is further suppressed by $\left(\alpha/{4\pi}\right)^2$ and does not give any meaningful constraint and is much below the sensitivity level achieved in the current or planned direct detection experiments~\cite{Kopp:2014tsa, Kopp:2009et, DEramo:2017zqw}. 
\begin{figure*}[!ht]
  \centering
  \subfloat[]{\includegraphics[width=0.48\linewidth]{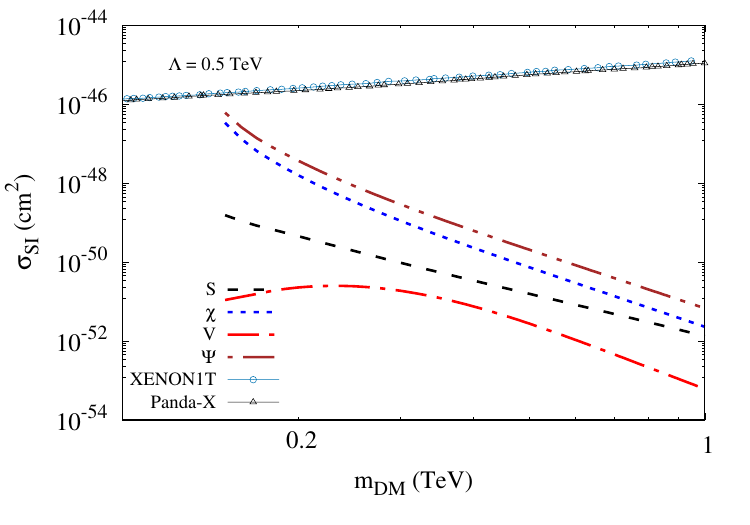}\label{fig:dd500}}
  \hfill
  \subfloat[]{\includegraphics[width=0.48\linewidth]{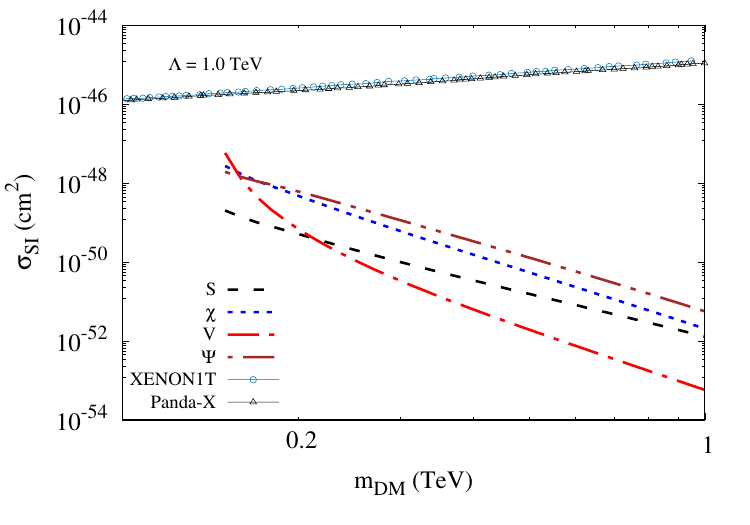}\label{fig:dd1000}}
  \\
  \subfloat[]{\includegraphics[width=0.48\linewidth]{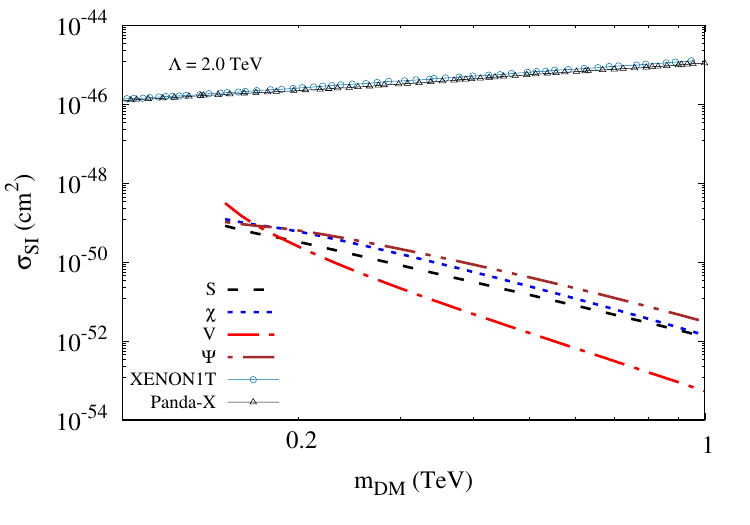}\label{fig:dd2000}}
  \hfill
  \subfloat[]{\includegraphics[width=0.48\linewidth]{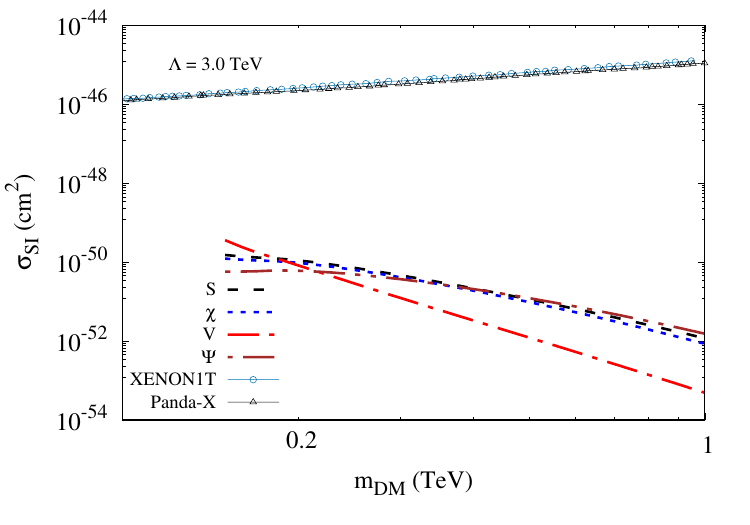}\label{fig:dd3000}}
  \caption{Dark matter-nucleon scattering cross-section as a function of DM mass in model {\bf A}. Current bounds on spin-independent interactions from experiments like PANDA 2X-II 2017~\cite{Cui:2017nnn} and XENON1T~\cite{Aprile:2015uzo,Aprile:2017aty} are also shown.  All points on the contour are consistent with the observed relic density $\Omega_{\rm DM} h^2 = 0.119$. The panels~(\ref{fig:dd500}~-~\ref{fig:dd3000})  correspond to the KK-graviton interaction scale  $\Lambda = 500$~GeV, 1~TeV, 2~TeV and 3~TeV respectively.}
  \label{fig:dd}
\end{figure*}

Using the expression~\cref{sigmasi} we have plotted the spin-independent DM-Nucleon scattering cross-section in the bench-mark model {\bf A}  as a function of DM mass in~\cref{fig:dd}. The panels~(\ref{fig:dd500}~-~\ref{fig:dd3000})  correspond to the KK-graviton interaction scale  $\Lambda = 500$~GeV, 1~TeV, 2~TeV and 3~TeV respectively. The parameter set used in the computation are consistent with the observed relic density given in~\cref{fig:relicA,fig:relicB}.  The upper limits from PANDA~2x-II 2017~\cite{Cui:2017nnn} and XENON-1T~\cite{Aprile:2015uzo,Aprile:2017aty} are also shown with the forbidden region.
\begin{figure*}[!ht]
  \centering
  \subfloat[]{\includegraphics[width=0.48\linewidth]{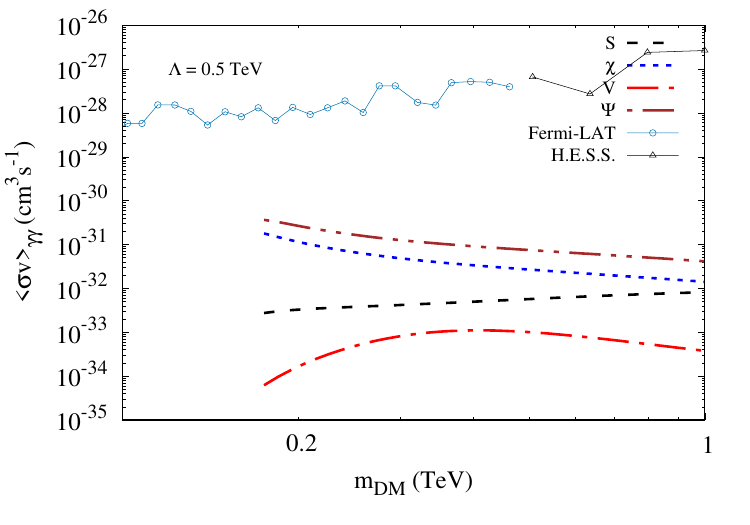}\label{fig:indirectA500}}
  \hfill
  \subfloat[]{\includegraphics[width=0.48\linewidth]{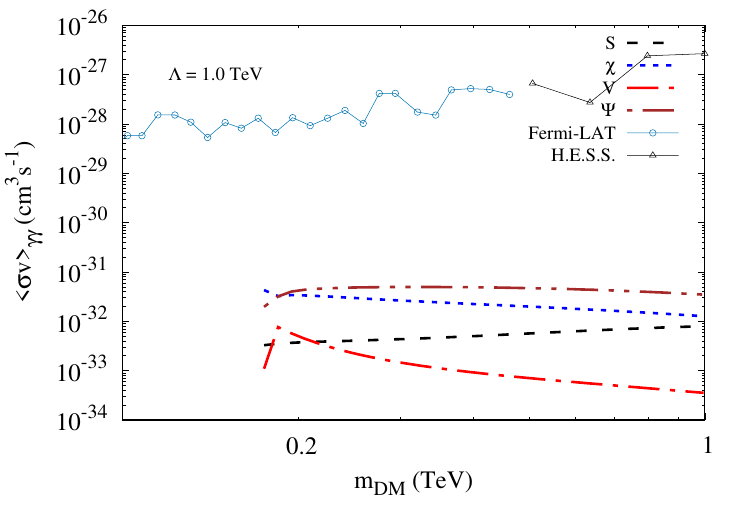}\label{fig:indirectA1000}}
  \\
  \subfloat[]{\includegraphics[width=0.48\linewidth]{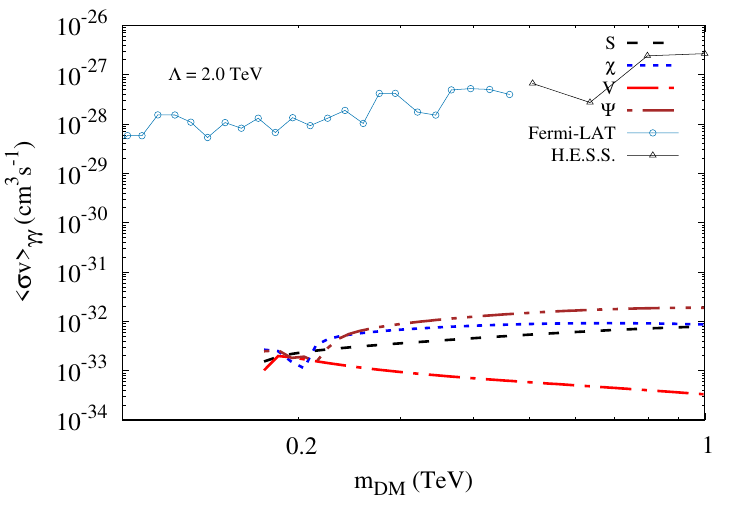}\label{fig:indirectA2000}}
  \hfill
  \subfloat[]{\includegraphics[width=0.48\linewidth]{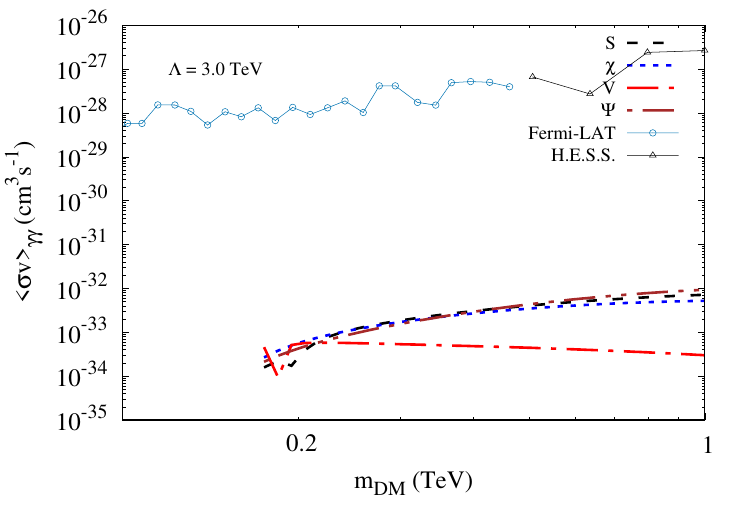}\label{fig:indirectA3000}}
  \caption{Velocity-averaged cross-section $\langle\sigma v\rangle_{\gamma \gamma}$ in the benchmark model {\bf A} . The current upper bounds from the Fermi-LAT~\cite{Ackermann:2015lka} and H.E.S.S~\cite{Abramowski:2013ax,Abdalla:2016olq} data are shown. All points on the contour satisfy the observed relic density. The panels (\ref{fig:indirectA500}~-~\ref{fig:indirectA3000}) correspond to the KK-graviton interaction scale  $\Lambda = 500$~GeV, 1~TeV, 2~TeV and 3~TeV respectively.}
  \label{fig:indirectA}
\end{figure*}
\begin{figure*}[!ht]
  \centering
  \subfloat[]{\includegraphics[width=0.48\linewidth]{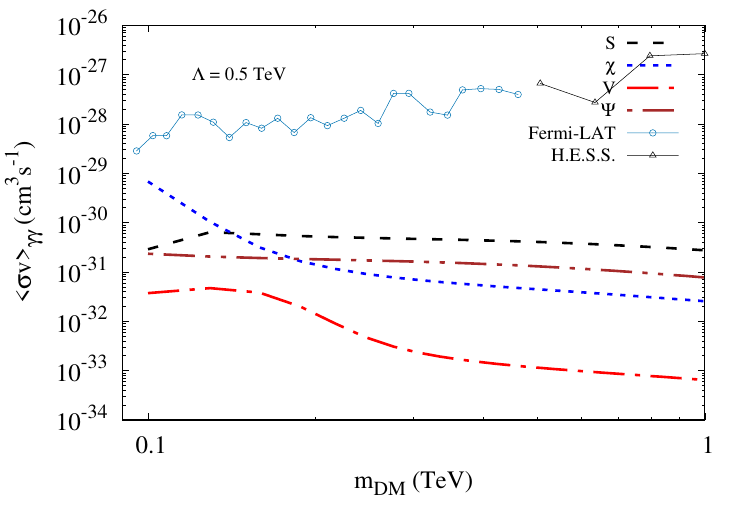}\label{fig:indirectB500}}
  \hfill
  \subfloat[]{\includegraphics[width=0.48\linewidth]{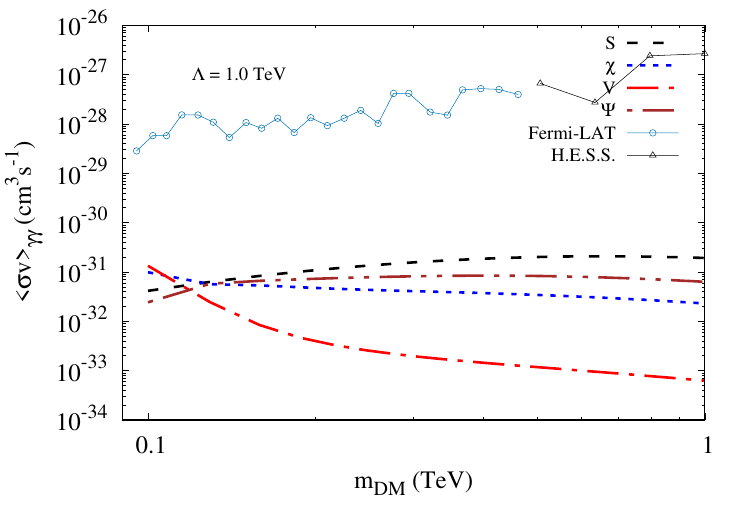}\label{fig:indirectB1000}}
  \\
  \subfloat[]{\includegraphics[width=0.48\linewidth]{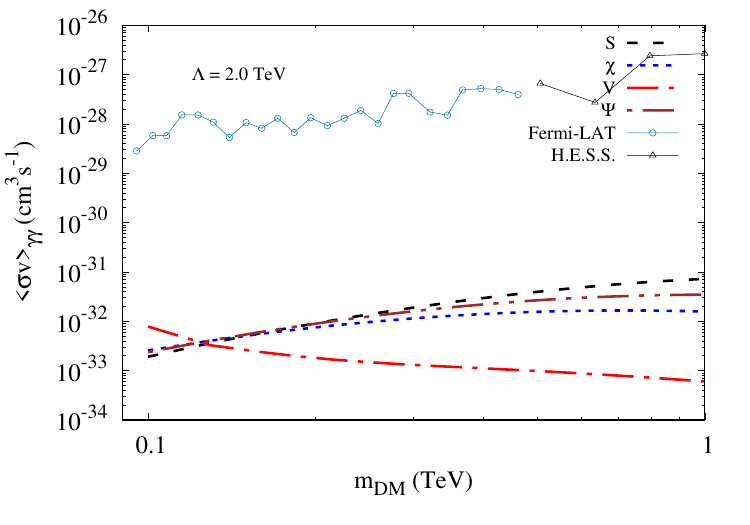}\label{fig:indirectB2000}}
  \hfill
  \subfloat[]{\includegraphics[width=0.48\linewidth]{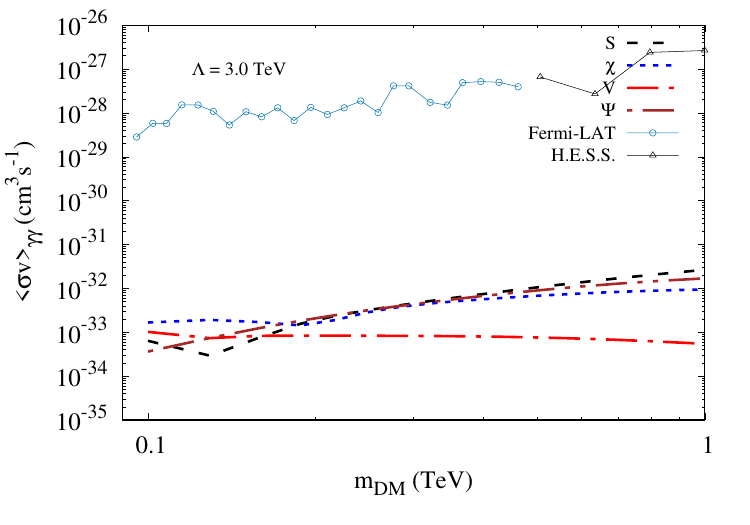}\label{fig:indirectB3000}}
  \caption{Velocity-averaged cross-section $\langle\sigma v\rangle_{\gamma \gamma}$ in the benchmark model {\bf B}. The current upper bounds from the Fermi-LAT~\cite{Ackermann:2015zua} data are shown. All points on the contour satisfy the observed relic density. The panels (\ref{fig:indirectB500}~-~\ref{fig:indirectB3000}) correspond to the KK-graviton interaction scale  $\Lambda = 500$~GeV, 1~TeV, 2~TeV and 3~TeV respectively.}
  \label{fig:indirectB}
\end{figure*}

\subsection{Indirect detection}
\label{sec:id}
Observation of diffused gamma rays from the regions of the galaxy such as Galactic centre (GC) and dwarf spheroidal galaxies (dSphs) where DM density appears to be large, imposes bounds on the DM annihilation to the SM particles. Fermi-LAT~\cite{Ackermann:2015lka, Ackermann:2015zua} and H.E.S.S~\cite{Abramowski:2013ax,Abdalla:2016olq} have investigated DM annihilation as a possible source of incoming photon flux. These experiments are then used to put constraints on the upper limit of velocity averaged scattering cross-sections for various channels which can contribute to the observed photon flux. The $\gamma\gamma$ photon line search in the photon spectrum in the KK graviton mediated DM annihilation into two photons occur directly as well as through the top-quark triangle loop for the model {\bf A} and charged lepton loop for model {\bf B} as discussed in \cref{sec:model}. In the case of lepto-philic DM model {\bf B}, the diffused $\gamma$-ray flux arises from the annihilation channel into charged leptons. The velocity averaged annihilation cross-section $\langle\sigma v\rangle_{\ell^+\ell^-}$ is, however, several orders of magnitude smaller than the corresponding $\gamma \gamma$ annihilation cross-section and will not be considered further.

The DM is gravitationally bound and is moving typically with velocity roughly lying between 200-500~$\rm km/s$ and the constraints obtained from indirect searches are not sensitive to the specific choice of the velocity. For the case of vector and spin-3/2 DMs, the choice of specific velocity is especially unimportant and $v\to 0$ is an excellent approximation. For our numerical calculation, we have kept the exact expression with the choice of $v= 300\, \rm km/s$. The experimental limits on this mode are given by the Fermi-LAT~\cite{Ackermann:2015lka} and H.E.S.S~\cite{Abramowski:2013ax,Abdalla:2016olq} Galactic Centre data sets. The limits depend upon the velocity distribution profile.

In~\cref{fig:indirectA} and~\cref{fig:indirectB}  we have plotted the variation of  velocity-averaged scattering cross-section $\langle\sigma v\rangle_{\gamma \gamma}$ with the DM mass in bench-mark model {\bf A} and {\bf B} respectively. We have also shown the observational constraints for the two photon final state annihilation rates in~\cref{fig:indirectA} and~\cref{fig:indirectB}. All parameters are chosen to satisfy the observed relic density. The panels~(\ref{fig:indirectA500}~-~\ref{fig:indirectA3000}) and~(\ref{fig:indirectB500}~-~\ref{fig:indirectB3000}) correspond to the graviton interaction scale $\Lambda = 500$~GeV, 1~TeV, 2~TeV and 3~TeV respectively for the benchmark models {\bf A} and {\bf B} respectively. We find that the annihilation cross-sections for the above processes in the benchmark models {\bf A} and {\bf B} are roughly three to four orders of magnitude smaller than the current upper bounds from the Fermi-LAT~\cite{Ackermann:2015lka, Ackermann:2015zua} and H.E.S.S~\cite{Abramowski:2013ax,Abdalla:2016olq} data.

\begin{figure*}[!ht]
  \centering
  \subfloat[]{\includegraphics[width=0.48\linewidth]{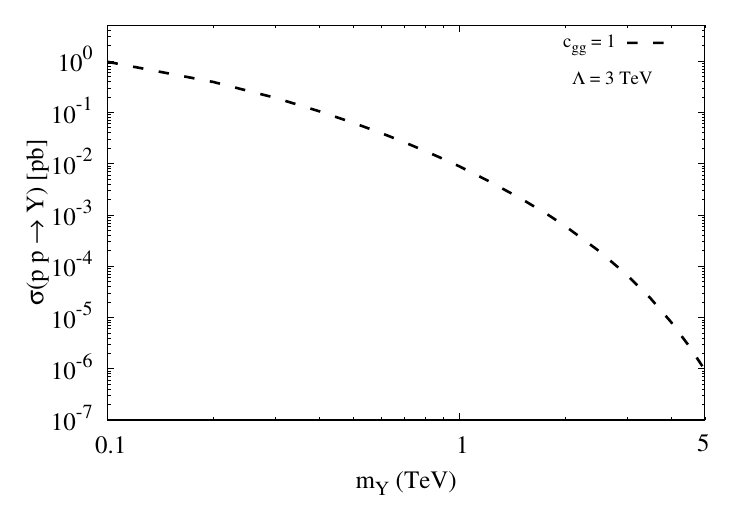}\label{fig:xsec}}
  \hfill
  \subfloat[]{\includegraphics[width=0.48\linewidth]{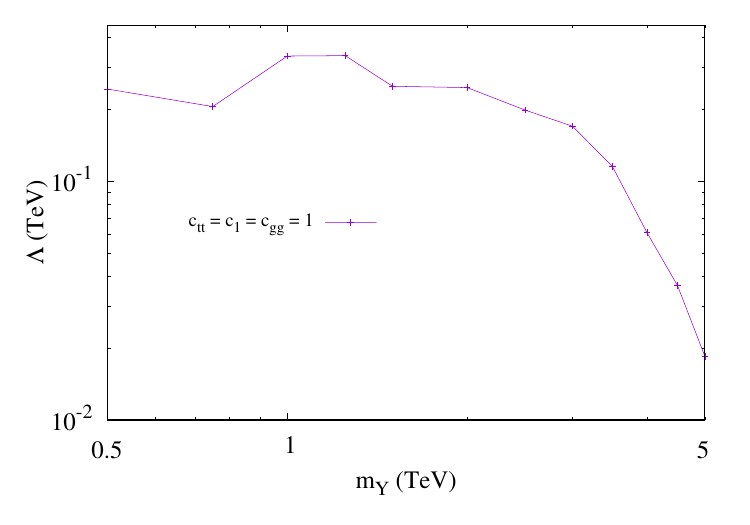}\label{fig:limit}}
  \caption{(a) KK graviton production LO cross-section multiplied by the K factor 1.3 through gluon fusion including the right-handed top loop. The cross-section is given for ${\sqrt s} = 13$~TeV for $\Lambda = 3$~TeV and the couplings $c_{tt} = c_{gg} =1$. The cross-section for different values of $\Lambda$ and the couplings can be obtained by scaling (see text). (b) Constraints on the KK graviton interaction scale $\Lambda$ as a function of graviton mass from the observed 95\% C.L. lower limits of the resonance searches for ${\sqrt s} = 13$~TeV at 35.9~fb$^{-1}$ by CMS ~\cite{Sirunyan:2018ryr}.}
  \label{fig:xsecl}
\end{figure*}

\subsection{LHC constraints}
\label{sec:lhc}
The s-channel graviton ($Y$) production $p p \to Y$ at LHC in our benchmark model {\bf A} is dominated by gluon fusion and through the right-handed top quark loop. We calculate the $Y$ production leading order (LO) cross-section at ${\sqrt s} =13$~TeV at 35.9 fb$^{-1}$ . In~\cref{fig:xsec}, $Y$ production LO cross-section multiplied by the K-factor 1.3~\cite{Bonciani:2015hgv} is shown for the graviton mass $m_Y$ lying between 100~GeV to 5~TeV by using the effective coupling $c_{gg}^{\rm eff}$. The interaction scale $\Lambda$ is taken to be 3~TeV and the coupling $c_{gg}^{\rm eff}$ is obtained by taking $c_{tt} = c_{gg} =1$ using expressions given in~\cref{eq:ceffGGA}-(\ref{eq:ceffggA}). As remarked earlier, the results are very weakly dependent on the couplings $c_{gg}$ and $c_{tt}$, even with a decrease by even two orders of magnitude, the production cross-section changes by roughly a factor of four because of relatively large contribution coming from the top quark loop. The cross-section for a given value of $\Lambda$ can be obtained by scaling.

In order to estimate the constraints on the graviton interaction scale $\Lambda$, we employ the recent CMS~\cite{Sirunyan:2018ryr} data for resonance searches in the narrow-width approximation for $t \bar t$ production cross-section at ${\sqrt s} = 13$~TeV with a luminosity of 35.9~fb$^{-1}$ and taking the $Y \to t \bar t$ branching ratio to be one. The $t \bar t$ final state of the KK graviton gives the strongest constraint on the scale factor $\Lambda$. In~\cref{fig:limit}, we have shown the constraints on the interaction scale $\Lambda$ from the observed 95\% C.L. lower limits given by CMS. We find that the lower limit on $\Lambda$ for all mass points $m_Y$ lying between 500~GeV to 5~TeV for which the data exists is less than a few hundred~GeV.

\section{Results and conclusions}
\label{sec:rc}
In this article, we have investigated the viability of a scalar, vector, a spin-1/2 or a spin-3/2 fermionic dark matter particle interacting purely gravitationally with the standard model particles mediated by its Kaluza-Klein graviton in the Randall-Sundram framework. Since the RS model with universal coupling to all SM particles is severally constrained in its parameter space by the LHC data~\cite{Kraml:2017atm}, we are led to consider the RS model with non-universal couplings to SM particles. In the top-philic model {\bf A} considered here, only the right-handed top quarks are assumed to interact strongly with the gravitons. The colour and $U(1)$ gauge bosons are assumed to live in the bulk. In the second benchmark lepto-philic model {\bf B} only the right-handed charged leptons interact strongly with the gravitons. The dark matter particles of any spin, on the other hand, are assumed to live on the IR brane and thus interact strongly with the gravitons.

We can see from~\cref{fig:relicA,fig:relicB} that in both these models there exists a large parameter space in which the observed relic density can be obtained for reasonable values of the parameters.
In~\cref{fig:dd} we have plotted the dark matter-nucleon scattering cross-section as a function of DM mass in model {\bf A}. Current bounds on spin-independent interactions from experiments like PANDA 2X-II 2017~\cite{Cui:2017nnn} and XENON1T~\cite{Aprile:2015uzo,Aprile:2017aty} are also shown.  Both the models {\bf A} and {\bf B} predict dark matter-nucleon scattering cross-sections much below the sensitivity levels achieved in the current or planned direct detection experiments. Another notable feature of the model {\bf A} is that the leading term in the DM-Nucleon cross-section is independent of the spin of the dark matter particle.

From the velocity-averaged cross-section $\langle\sigma v\rangle_{\gamma \gamma}$ in the benchmark model {\bf A} (\cref{fig:indirectA}) and model {\bf B} (\cref{fig:indirectB}), we find that the annihilation cross-sections for the above processes are roughly three to four orders of magnitude smaller than the current upper bounds from the Fermi-LAT~\cite{Ackermann:2015lka, Ackermann:2015zua} and H.E.S.S~\cite{Abramowski:2013ax,Abdalla:2016olq} data.

In conclusion, the top-philic as well as the lepto-philic models in the Randall-Sundram framework discussed in this work, are capable of explaining the observed relic density for a reasonable set of parameters without transgressing the constraints from the direct and indirect experiments. They are also consistent with the LHC data~\cite{Kraml:2017atm}.
They are also consistent with the recent CMS~\cite{Sirunyan:2018ryr} data for resonance searches in the narrow-width approximation for $t \bar t$ production cross-section at ${\sqrt s} = 13$~TeV with a luminosity of 35.9~fb$^{-1}$ for $m_Y$ lying between 500~GeV to 5~TeV for which the data exists, albeit for the graviton mass roughly greater than 500~GeV with the truncated dark matter mass range required to obtain the requisite relic density.

\acknowledgments
AG and RI would like to thank the SA-CERN Consortium and Institute for Collider Particle Physics, University of the Witwatersrand for their support and hospitality during their visit. RI would also like to thank the SERB-DST, India for the research grant EMR/2015/000333 and Rishav Roshan and Sreemanti Chakraborti for useful discussions.

\appendix
\section{KK graviton-gauge boson effective couplings and decay widths}
\label{sec:a}
The effective KK graviton-gauge boson couplings in benchmark models {\bf A} and {\bf B} are
\begin{description}[leftmargin=0pt,labelindent=0pt]
  \item[Model A]
  \begin{gather}
    c^{\rm eff}_{gg}
    =
    c_{gg} (m_Y) + c_{tt}\,F^t_Y \left(4\frac{m^2_t}{m^2_Y}\right)
    \label{eq:ceffGGA}
    \\
    c^{\rm eff}_{\gamma\gamma}
    =
    c_{\gamma\gamma} (m_Y) + 2 Q^2_t\,N_c\,c_{tt}\,F^t_Y \left(4\frac{m^2_t}{m^2_Y}\right)
    \label{eq:ceffggA}
    \\
    F^t_Y \left(4\frac{m^2_t}{m^2_Y}\right)
    =
    \begin{cases}
      A_Y \left(4\frac{m^2_t}{m^2_Y}\right), & m_Y \geq m_t \\
      B_Y \left(4\frac{m^2_t}{m^2_Y}\right), & m_Y < m_t
    \end{cases}
  \end{gather}
   where \(Q_t = 2/3\) and \(N_c = 3\).

  \item[Model B]
  \begin{gather}
    c^{\rm eff}_{\gamma\gamma}
    =
    c_{\gamma\gamma} (m_Y) + \sum_{\ell=e,\mu,\tau} 2 Q^2_{\ell}\,c_{\ell\ell}\,F^{\ell}_Y \left(4\frac{m^2_{\ell}}{m^2_Y}\right)
    \label{eq:ceffggB}
    \\
    F^{\ell}_Y \left(4\frac{m^2_{\ell}}{m^2_Y}\right)
    =
    A_Y \left(4\frac{m^2_{\ell}}{m^2_Y}\right)
  \end{gather}
  and \(Q_\ell = -1, m_\ell = m_e, m_\mu, m_\tau \ll m_Y\).
  \begin{align}
    A_Y(\tau)
    =&\,
    - \frac{1}{36} \left[9\tau \left( \tau+2\right) f(\tau) + 6\left( 5\tau+4\right) g(\tau) - 39\tau-35 + 12\, {\rm ln}(\tau/4) \right], & \tau \le 4 \\
    B_Y(\tau)
    =&\,
    - \frac{1}{36} \left[9\tau \left( \tau+2\right) f(\tau) + 6\left( 5\tau+4\right) g(\tau) - 39\tau-35  \right], & \tau > 4 \\
    f(\tau)
    =&\,
    \begin{cases}
      - \left[\tanh^{-1}\left( \sqrt{1-\tau}\right) - i \pi/2 \right]^2, & \qquad \tau < 1 \\
      \left[\sin^{-1}\left( 1/\sqrt{\tau}\right)\right]^2,               & \qquad \tau \geq 1
    \end{cases}
    \\
    g(\tau)
    =&\,
    \begin{cases}
      \sqrt{1-\tau}\left[\tanh^{-1}\left( \sqrt{1-\tau}\right) - i \pi/2  \right], & \tau < 1 \\
      \sqrt{\tau-1}\, \sin^{-1}\left( 1/\sqrt{\tau}\right),                        & \tau \ge 1
    \end{cases}
  \end{align}
\end{description}
KK graviton decay widths relevant for the benchmark models A and B are given as follows
\begin{align}
  \Gamma(Y \to \ell\bar\ell)
  =& 
  \frac{m^3_Y}{320 \pi \Lambda^2} c^2_{\ell\ell} ,
  \\%\nl
  \Gamma(Y \to t\bar t)
  =& 
  \frac{N_c m^3_Y}{320 \pi \Lambda^2} \left( 1 - 4 \frac{m^2_t}{m^2_Y} \right)^{\sfrac{3}{2}} \left( 1 - \frac{2}{3} \frac{m^2_t}{m^2_Y} \right) c^2_{tt} ,
  \\%\nl
  \Gamma(Y \to gg)
  =& 
  \frac{m^3_Y}{10 \pi \Lambda^2} \left|\frac{\alpha_S}{4\pi} c^{\rm eff}_{gg}\right|^2 ,
  \\%\nl
  \Gamma(Y \to \gamma\gamma)
  =& 
  \frac{m^3_Y}{80 \pi \Lambda^2} \left|\frac{\alpha}{4\pi} c^{\rm eff}_{\gamma\gamma}\right|^2 ,
%  \\
\end{align}
\begin{align}
  \Gamma(Y \to \gamma Z)
  =& 
  \frac{m^3_Y}{40 \pi \Lambda^2} \left( 1 - \frac{m^2_Z}{m^2_Y} \right)^3 \left( 1 + \frac{1}{2} \frac{m^2_Z}{m^2_Y} + \frac{1}{6} \frac{m^4_Z}{m^4_Y} \right) \left|\frac{\alpha}{4\pi} c^{\rm eff}_{\gamma Z}\right|^2 ,
  \\%\nl
  \Gamma(Y \to ZZ)
  =& 
  \frac{m^3_Y}{80 \pi \Lambda^2} \left( 1 - 4 \frac{m^2_Z}{m^2_Y} \right)^{\sfrac{1}{2}} \left( 1 - 3 \frac{m^2_Z}{m^2_Y} + 6 \frac{m^4_Z}{m^4_Y} \right) \left|\frac{\alpha}{4\pi} c^{\rm eff}_{ZZ}\right|^2 .
\end{align}
The KK graviton decay widths into DM particle pairs are
\begin{align}
  \Gamma(Y \to SS)
  =& 
  \frac{m^3_Y}{960 \pi \Lambda^2} \left( 1 - 4 \frac{m^2_S}{m^2_Y} \right)^{\sfrac{5}{2}} c^2_{SS} ,
  \\%\nl
  \Gamma(Y \to \chi\bar\chi)
  =& 
  \frac{m^3_Y}{480 \pi \Lambda^2} \left( 1 - 4 \frac{m^2_\chi}{m^2_Y} \right)^{\sfrac{3}{2}} \left( 1 + \frac{8}{3} \frac{m^2_\chi}{m^2_Y} \right) c^2_{\chi\chi} ,
\\%\nl
  \Gamma(Y \to VV)
  =& 
  \frac{m^3_Y}{960 \pi \Lambda^2} \left( 1 - 4 \frac{m^2_V}{m^2_Y} \right)^{\sfrac{1}{2}} \left( 13 + 56 \frac{m^2_V}{m^2_Y} + 48 \frac{m^4_V}{m^4_Y}\right) c^2_{VV} ,
\\%\nl
  \Gamma(Y \to \Psi\bar\Psi)
  =& 
  \frac{m^7_Y}{1440 \pi \Lambda^2 m_\Psi^4} \left( 1 - 4 \frac{m^2_\Psi}{m^2_Y} \right)^{\sfrac{3}{2}} \left( 1 - \frac{4}{3} \frac{m^2_\Psi}{m^2_Y} - 6 \frac{m^4_\Psi}{m^4_Y} + 48 \frac{m^6_\Psi}{m^6_Y} \right) c^2_{\Psi\Psi} . \label{decay-kk}
\end{align}
Here, $S, \chi, V$ and $\Psi$ are the spin-0, spin-1/2, spin-1 and spin-3/2 DM particles respectively.

\section{Annihilation cross-sections}
\label{sec:b}
The annihilation cross-sections of DM into various final states relevant for the benchmark models A and B are given as follows:

\subsection{Scalar dark matter}
\begin{align}
  \sigma_{SS \to \ell^+\ell^-}
  &=
  \frac{1}{32\pi s}\!\!
  \left[\frac{s}{s - 4m^2_S}\right]^{\sfrac{1}{2}}\nquad
  \frac{1}{[(s-m_Y^2)^2+\Gamma_Y^2 m_Y^2]}
  \frac{c_{SS}^2 c_{\ell\ell}^2\,s^2}{60\Lambda^4}
  \left( s - 4 m_S^2 \right)^2
  \\
  \sigma_{SS \to t\bar t}
  =&
  \frac{1}{32\pi s}\!\!
  \left[\frac{s - 4m^2_t}{s - 4m^2_S}\right]^{\sfrac{1}{2}}\nquad
  \frac{1}{[(s-m_Y^2)^2+\Gamma_Y^2 m_Y^2]}
  \frac{c_{SS}^2 c_{tt}^2}{60\Lambda^4}
  \left( s-4 m_t^2 \right)
  \nl
  &\times
  \Big[
    \left(3 s-2 m_t^2\right) \left(s-4 m_S^2\right)^2
    + 20 m_t^2 \Big(1-\frac{s}{m_Y^2}\Big)^2 \left(2 m_S^2+s\right)^2
  \Big]
  \\
  \sigma_{SS \to gg}
  =&
  \frac{1}{32\pi s}\!\!
  \left[\frac{s}{s - 4m^2_S}\right]^{\sfrac{1}{2}}\nquad
  \frac{1}{[(s-m_Y^2)^2+\Gamma_Y^2 m_Y^2]}
  \frac{8 c_{SS}^2\,s^2}{45\Lambda^4}
  \left|\frac{\alpha_S}{4\pi} c^{\rm eff}_{gg}\right|^2
  \nl
  &\times
  \Big[
    3 \left(s-4 m_S^2\right)^2
    + 10 \Big(1-\frac{s}{m_Y^2}\Big)^2 \left(2 m_S^2+s\right)^2
  \Big]
  \\
  \sigma_{SS \to \gamma\gamma}
  =&
  \frac{1}{32\pi s}\!\!
  \left[\frac{s}{s - 4m^2_S}\right]^{\sfrac{1}{2}}\nquad
  \frac{1}{[(s-m_Y^2)^2+\Gamma_Y^2 m_Y^2]}
  \frac{c_{SS}^2\,s^2}{45\Lambda^4}
  \left|\frac{\alpha}{4\pi} c^{\rm eff}_{\gamma\gamma}\right|^2
  \nl
  &\times
  \Big[
    3 \left(s-4 m_S^2\right)^2
    + 10 \Big(1-\frac{s}{m_Y^2}\Big)^2 \left(2 m_S^2+s\right)^2
  \Big]
%  \\
\end{align}
\begin{align}
  \sigma_{SS \to \gamma Z}
  =&
  \frac{1}{32\pi s}
  \frac{s - m^2_Z}{\left[s(s - 4m^2_S)\right]^{\sfrac{1}{2}}}
  \frac{1}{[(s-m_Y^2)^2+\Gamma_Y^2 m_Y^2]}
  \frac{c_{SS}^2}{45\Lambda^4}
  \left|\frac{\alpha}{4\pi} c^{\rm eff}_{\gamma Z}\right|^2
  \frac{\left( s - m^2_Z \right)^2\!\!\!}{s^2}
  \nl
  \times&
  \Big[
    \left(s-4 m_S^2\right)^2\!\!\left(m_Z^4+3 m_Z^2 s+6 s^2\right)
    + 5 \Big(1-\frac{s}{m_Y^2}\Big)^2\!\!\left(2 m_S^2+s\right)^2\!\!\left(m_Z^2+2 s\right)^2
  \Big]
  \\
 \sigma_{SS \to ZZ}
  =&
  \frac{1}{32\pi s}\!\!
  \left[\frac{s - 4m^2_Z}{s - 4m^2_S}\right]^{\sfrac{1}{2}}\nquad
  \frac{1}{[(s-m_Y^2)^2+\Gamma_Y^2 m_Y^2]}
  \frac{c_{SS}^2}{45\Lambda^4}
  \left|\frac{\alpha}{4\pi} c^{\rm eff}_{ZZ}\right|^2
  \Big[
    3 \left(s-4 m_S^2\right)^2
  \nl
  \times&
    \left(6 m_Z^4-3 m_Z^2 s+s^2\right)
    + 5 \Big(1-\frac{s}{m_Y^2}\Big)^2\!\!\left(2 m_S^2+s\right)^2 \left(11 m_Z^4-4 m_Z^2 s+2 s^2\right)
  \Big]
\end{align}

\subsection{Vector dark matter}
\begin{align}
  \sigma_{VV \to \ell^+\ell^-}\!\!
  =&
  \frac{1}{32\pi s}\!\!
  \left[\!\frac{s}{s - 4m^2_V}\!\right]^{\sfrac{1}{2}}\nquad
  \frac{1}{[(s-m_Y^2)^2+\Gamma_Y^2 m_Y^2]}
  \frac{c_{VV}^2 c_{\ell\ell}^2\,s^2}{540\Lambda^4}
  \Big[
    48 m_V^4+56 m_V^2 s+13 s^2
  \Big]
  \\
  \sigma_{VV \to t\bar t}
  =&
  \frac{1}{32\pi s}\!\!
  \left[\frac{s - 4m^2_t}{s - 4m^2_V}\right]^{\sfrac{1}{2}}\nquad
  \frac{1}{[(s-m_Y^2)^2+\Gamma_Y^2 m_Y^2]}
  \frac{c_{VV}^2 c_{tt}^2}{540\Lambda^4}
  \left( s - 4 m_t^2 \right)
  \Big[
    \left(3 s-2 m_t^2\right)
  \nl
  \times&
    \left(48 m_V^4+56 m_V^2 s+13 s^2\right)
    + 20 m_t^2 \Big(1-\frac{s}{m_Y^2}\Big)^2 \left(12 m_V^4-4 m_V^2 s+s^2\right)
  \Big]
  \\
  \sigma_{VV \to gg}
  =&
  \frac{1}{32\pi s}\!\!
  \left[\frac{s}{s - 4m^2_V}\right]^{\sfrac{1}{2}}\nquad
  \frac{1}{[(s-m_Y^2)^2+\Gamma_Y^2 m_Y^2]}
  \frac{8 c_{VV}^2\,s^2}{405\Lambda^4}
  \left|\frac{\alpha_S}{4\pi} c^{\rm eff}_{gg}\right|^2
  \nl
  \times&
  \Big[
    3 \left(48 m_V^4+56 m_V^2 s+13 s^2\right)
    + 10 \Big(1-\frac{s}{m_Y^2}\Big)^2 \left(12 m_V^4-4 m_V^2 s+s^2\right)
  \Big]
  \\
  \sigma_{VV \to \gamma\gamma}
  =&
  \frac{1}{32\pi s}\!\!
  \left[\frac{s}{s - 4m^2_V}\right]^{\sfrac{1}{2}}\nquad
  \frac{1}{[(s-m_Y^2)^2+\Gamma_Y^2 m_Y^2]}
  \frac{c_{VV}^2\,s^2}{405\Lambda^4}
  \left|\frac{\alpha}{4\pi} c^{\rm eff}_{\gamma\gamma}\right|^2
  \nl
  \times&
  \Big[
    3 \left(48 m_V^4+56 m_V^2 s+13 s^2\right)
    + 10 \Big(1-\frac{s}{m_Y^2}\Big)^2 \left(12 m_V^4-4 m_V^2 s+s^2\right)
  \Big]
  \\
  \sigma_{VV \to \gamma Z}
  =&
  \frac{1}{32\pi s}
  \frac{s - m^2_Z}{\left[s(s - 4m^2_V)\right]^{\sfrac{1}{2}}}
  \frac{1}{[(s-m_Y^2)^2+\Gamma_Y^2 m_Y^2]}
  \frac{c_{VV}^2}{405\Lambda^4}
  \left|\frac{\alpha}{4\pi} c^{\rm eff}_{\gamma Z}\right|^2
  \frac{( s - m_Z^2 )^2\!\!}{s^2}
  \nl
  &\times
  \Big[
    \left(48 m_V^4+56 m_V^2 s+13 s^2\right) \left(m_Z^4+3 m_Z^2 s+6 s^2\right)
  \nl
  &
    + 5 \Big(1-\frac{s}{m_Y^2}\Big)^2 \left(12 m_V^4-4 m_V^2 s+s^2\right) \left(m_Z^2+2 s\right)^2
  \Big]
  \\
  \sigma_{VV \to ZZ}
  =&
  \frac{1}{32\pi s}\!\!
  \left[\frac{s - 4m^2_Z}{s - 4m^2_V}\right]^{\sfrac{1}{2}}\nquad
  \frac{1}{[(s-m_Y^2)^2+\Gamma_Y^2 m_Y^2]}
  \frac{c_{VV}^2}{405\Lambda^4}
  \left|\frac{\alpha}{4\pi} c^{\rm eff}_{ZZ}\right|^2
  \nl
  &\times
  \Big[
    3 \left(48 m_V^4+56 m_V^2 s+13 s^2\right) \left(6 m_Z^4-3 m_Z^2 s+s^2\right)
  \nl
  &
    + 5 \Big(1-\frac{s}{m_Y^2}\Big)^2 \left(12 m_V^4-4 m_V^2 s+s^2\right) \left(11 m_Z^4-4 m_Z^2 s+2 s^2\right)
  \Big]
\end{align}

\subsection{Spin-1/2 dark matter}
\begin{align}
  \sigma_{\chi\bar\chi \to \ell^+\ell^-}
  =&
  \frac{1}{32\pi s}\!\!
  \left[\frac{s}{s - 4m^2_\chi}\right]^{\sfrac{1}{2}}\nquad
  \frac{1}{[(s-m_Y^2)^2+\Gamma_Y^2 m_Y^2]}
  \frac{c_{\chi\chi}^2 c_{\ell\ell}^2\,s^2}{240\Lambda^4}
  \left( s - 4 m_\chi^2 \right)
  \left( 8 m_\chi^2+3 s \right)
%  \\
\end{align}
\begin{align}
  \sigma_{\chi\bar\chi \to t\bar t}
  =&
  \frac{1}{32\pi s}\!\!
  \left[\frac{s - 4m^2_t}{s - 4m^2_\chi}\right]^{\sfrac{1}{2}}\nquad
  \frac{1}{[(s-m_Y^2)^2+\Gamma_Y^2 m_Y^2]}
  \frac{c_{\chi\chi}^2 c_{tt}^2}{240\Lambda^4}
  \left( s - 4 m_\chi^2 \right) \left( s - 4 m_t^2 \right)
  \nl
  &\times
  \Big[
    \left(3 s-2 m_t^2\right) \left(8 m_\chi^2+3 s\right)
    + 40 m_t^2 m_\chi^2 \Big(1-\frac{s}{m_Y^2}\Big)^2
  \Big]
  \\
  \sigma_{\chi\bar\chi \to gg}
  =&
  \frac{1}{32\pi s}\!\!
  \left[\frac{s}{s - 4m^2_\chi}\right]^{\sfrac{1}{2}}\nquad
  \frac{1}{[(s-m_Y^2)^2+\Gamma_Y^2 m_Y^2]}
  \frac{2 c_{\chi\chi}^2}{45\Lambda^4}
  \left|\frac{\alpha_S}{4\pi} c^{\rm eff}_{gg}\right|^2\!\!
  s^2 ( s - 4 m_\chi^2 )
  \nl
  &\times
  \Big[
    24 m_\chi^2+9 s
    + 20 m_\chi^2 \Big(1-\frac{s}{m_Y^2}\Big)^2
  \Big]
  \\
  \sigma_{\chi\bar\chi \to \gamma\gamma}
  =&
  \frac{1}{32\pi s}\!\!
  \left[\frac{s}{s - 4m^2_\chi}\right]^{\sfrac{1}{2}}\nquad
  \frac{1}{[(s-m_Y^2)^2+\Gamma_Y^2 m_Y^2]}
  \frac{c_{\chi\chi}^2}{180\Lambda^4}
  \left|\frac{\alpha}{4\pi} c^{\rm eff}_{\gamma\gamma}\right|^2\!\!
  s^2 ( s - 4 m_\chi^2 )
  \nl
  &\times
  \Big[
    24 m_\chi^2+9 s
    + 20 m_\chi^2 \Big(1-\frac{s}{m_Y^2}\Big)^2
  \Big]
  \\
  \sigma_{\chi\bar\chi \to \gamma Z}
  =&
  \frac{1}{32\pi s}
  \frac{s - m^2_Z}{\left[s(s - 4m^2_\chi)\right]^{\sfrac{1}{2}}}
  \frac{1}{[(s-m_Y^2)^2+\Gamma_Y^2 m_Y^2]}
  \frac{c_{\chi\chi}^2}{180\Lambda^4}
  \left|\frac{\alpha}{4\pi} c^{\rm eff}_{\gamma Z}\right|^2\!\!
  ( s - 4 m_\chi^2 )
  \frac{( s - m_Z^2 )^2\!\!\!}{s^2}
  \nl
  &\times
  \Big[
    \left(8 m_\chi^2+3 s\right)\!\!\left(m_Z^4+3 m_Z^2 s+6 s^2\right)
    + 10 m_\chi^2 \Big(1-\frac{s}{m_Y^2}\Big)^2\!\!\left(m_Z^2+2 s\right)^2
  \Big]
  \\
  \sigma_{\chi\bar\chi \to ZZ}
  =&
  \frac{1}{32\pi s}\!\!
  \left[\frac{s - 4m^2_Z}{s - 4m^2_\chi}\right]^{\sfrac{1}{2}}\nquad
  \frac{1}{[(s-m_Y^2)^2+\Gamma_Y^2 m_Y^2]}
  \frac{c_{\chi\chi}^2}{180\Lambda^4}
  \left|\frac{\alpha}{4\pi} c^{\rm eff}_{ZZ}\right|^2\!\!
  \left( s - 4 m_\chi^2 \right)\!\!
  \Big[
    3 \left(8 m_\chi^2+3 s\right)
  \nl
  &\times
    \left(6 m_Z^4-3 m_Z^2 s+s^2\right)
    + 10 m_\chi^2 \Big(1-\frac{s}{m_Y^2}\Big)^2 \left(11 m_Z^4-4 m_Z^2 s+2 s^2\right)
  \Big]
\end{align}

\subsection{Spin-3/2 dark matter}
\begin{align}
  \sigma_{\Psi\bar\Psi \to \ell^+\ell^-}
  &=
  \frac{1}{32\pi s}\!\!
  \left[\frac{s}{s - 4m^2_\Psi}\right]^{\sfrac{1}{2}}\nquad
  \frac{1}{[(s-m_Y^2)^2+\Gamma_Y^2 m_Y^2]}
  \frac{c_{\Psi\Psi}^2 c_{\ell\ell}^2\,s^2}{8640\Lambda^4 m_\Psi^4}
  \left( s - 4 m_\Psi^2 \right)
  \nl
  &\times
  \Big[
    144 m_\Psi^6-18 m_\Psi^4 s-4 m_\Psi^2 s^2+3 s^3
  \Big]
  \\
  \sigma_{\Psi\bar\Psi \to t\bar t}
  =&
  \frac{1}{32\pi s}\!\!
  \left[\frac{s - 4m^2_t}{s - 4m^2_\Psi}\right]^{\sfrac{1}{2}}\nquad
  \frac{1}{[(s-m_Y^2)^2+\Gamma_Y^2 m_Y^2]}
  \frac{c_{\Psi\Psi}^2 c_{tt}^2}{8640\Lambda^4 m_\Psi^4}
  \left( s - 4 m_\Psi^2 \right) \left( s - 4 m_t^2 \right)
  \nl
  &\times
  \Big[
    \left(144 m_\Psi^6-18 m_\Psi^4 s-4 m_\Psi^2 s^2+3 s^3\right) \left(3 s-2 m_t^2\right)
  \nl
  &
    + 40 m_\Psi^2 m_t^2 \Big(1-\frac{s}{m_Y^2}\Big)^2 \left(18 m_\Psi^4-6 m_\Psi^2 s+s^2\right)
  \Big]
  \\
  \sigma_{\Psi\bar\Psi \to gg}
  =&
  \frac{1}{32\pi s}\!\!
  \left[\frac{s}{s - 4m^2_\Psi}\right]^{\sfrac{1}{2}}\nquad
  \frac{1}{[(s-m_Y^2)^2+\Gamma_Y^2 m_Y^2]}
  \frac{c_{\Psi\Psi}^2\,s^2}{810\Lambda^4 m_\Psi^4}
  \left|\frac{\alpha_S}{4\pi} c^{\rm eff}_{gg}\right|^2
  ( s - 4 m_\Psi^2 )
  \Big[
    432 m_\Psi^6
  \nl
  &
    -54 m_\Psi^4 s-12 m_\Psi^2 s^2+9 s^3
    + 20 m_\Psi^2 \Big(1-\frac{s}{m_Y^2}\Big)^2 \left(18 m_\Psi^4-6 m_\Psi^2 s+s^2\right)
  \Big]
  \\
  \sigma_{\Psi\bar\Psi \to \gamma\gamma}
  =&
  \frac{1}{32\pi s}\!\!
  \left[\frac{s}{s - 4m^2_\Psi}\right]^{\sfrac{1}{2}}\nquad
  \frac{1}{[(s-m_Y^2)^2+\Gamma_Y^2 m_Y^2]}
  \frac{c_{\Psi\Psi}^2\,s^2}{6480\Lambda^4 m_\Psi^4}
  \left|\frac{\alpha}{4\pi} c^{\rm eff}_{\gamma\gamma}\right|^2
  ( s - 4 m_\Psi^2 )
  \Big[
    432 m_\Psi^6
  \nl
  &
    -54 m_\Psi^4 s-12 m_\Psi^2 s^2+9 s^3
    + 20 m_\Psi^2 \Big(1-\frac{s}{m_Y^2}\Big)^2 \left(18 m_\Psi^4-6 m_\Psi^2 s+s^2\right)
  \Big]
%  \\
\end{align}
\begin{align}
  \sigma_{\Psi\bar\Psi \to \gamma Z}
  =&
  \frac{1}{32\pi s}
  \frac{s - m^2_Z}{\left[s(s - 4m^2_\Psi)\right]^{\sfrac{1}{2}}}
  \frac{1}{[(s-m_Y^2)^2+\Gamma_Y^2 m_Y^2]}
  \frac{c_{\Psi\Psi}^2}{6480\Lambda^4 m_\Psi^4}
  \left|\frac{\alpha}{4\pi} c^{\rm eff}_{\gamma Z}\right|^2
  ( s - 4 m_\Psi^2 )
  \nl
  &\times
  \frac{( s - m_Z^2 )^2}{s^2}
  \Big[
    \left(144 m_\Psi^6-18 m_\Psi^4 s-4 m_\Psi^2 s^2+3 s^3\right) \left(m_Z^4+3 m_Z^2 s+6 s^2\right)
  \nl
  &
    + 10 m_\Psi^2 \Big(1-\frac{s}{m_Y^2}\Big)^2 \left(18 m_\Psi^4-6 m_\Psi^2 s+s^2\right) \left(m_Z^2+2 s\right)^2
  \Big]
  \\
  \sigma_{\Psi\bar\Psi \to ZZ}
  =&
  \frac{1}{32\pi s}\!\!
  \left[\frac{s - 4m^2_Z}{s - 4m^2_\Psi}\right]^{\sfrac{1}{2}}\nquad
  \frac{1}{[(s-m_Y^2)^2+\Gamma_Y^2 m_Y^2]}
  \frac{c_{\Psi\Psi}^2}{6480\Lambda^4 m_\Psi^4}
  \left|\frac{\alpha}{4\pi} c^{\rm eff}_{ZZ}\right|^2
  ( s - 4 m_\Psi^2 )
  \nl
  &\times
  \Big[
    3 \left(144 m_\Psi^6-18 m_\Psi^4 s-4 m_\Psi^2 s^2+3 s^3\right) \left(6 m_Z^4-3 m_Z^2 s+s^2\right)
  \nl
  &
    + 10 m_\Psi^2 \Big(1-\frac{s}{m_Y^2}\Big)^2 \left(18 m_\Psi^4-6 m_\Psi^2 s+s^2\right) \left(11 m_Z^4-4 m_Z^2 s+2 s^2\right)
  \Big]
\end{align}
The effective coupling constants \(c^{\rm eff}_{gg}, c^{\rm eff}_{\gamma\gamma}, c^{\rm eff}_{\gamma Z}\) and \(c^{\rm eff}_{ZZ}\) can be obtained from \cref{eq:ceffGGA,eq:ceffggA} for benchmark model {\bf A} and from \cref{eq:ceffggB} for model {\bf B} respectively replacing the KK graviton mass \(m_Y\) with the centre of mass energy \(\sqrt{s}\).

%%-------------------------------------------------------------------------------------------------%%

% \nocite{*} %% Show all the entries of .bib file in the References
 \bibliography{biblio}

\providecommand{\href}[2]{#2}\begingroup\raggedright\begin{thebibliography}{10}

\bibitem{Aghanim:2018eyx}
{\scshape Planck} collaboration, N.~Aghanim et~al., \emph{{Planck 2018 results.
  VI. Cosmological parameters}},  \href{https://arxiv.org/abs/1807.06209}{{\tt
  1807.06209}}.

\bibitem{Agnes:2015ftt}
{\scshape DarkSide} collaboration, P.~Agnes et~al., \emph{{Results From the
  First Use of Low Radioactivity Argon in a Dark Matter Search}},
  \href{http://dx.doi.org/10.1103/PhysRevD.93.081101,
  10.1103/PhysRevD.95.069901}{\emph{Phys. Rev.} {\bf D93} (2016) 081101},
  [\href{https://arxiv.org/abs/1510.00702}{{\tt 1510.00702}}].

\bibitem{Akerib:2016vxi}
{\scshape LUX} collaboration, D.~S. Akerib et~al., \emph{{Results from a search
  for dark matter in the complete LUX exposure}},
  \href{http://dx.doi.org/10.1103/PhysRevLett.118.021303}{\emph{Phys. Rev.
  Lett.} {\bf 118} (2017) 021303},
  [\href{https://arxiv.org/abs/1608.07648}{{\tt 1608.07648}}].

\bibitem{Aprile:2017iyp}
{\scshape XENON} collaboration, E.~Aprile et~al., \emph{{First Dark Matter
  Search Results from the XENON1T Experiment}},
  \href{http://dx.doi.org/10.1103/PhysRevLett.119.181301}{\emph{Phys. Rev.
  Lett.} {\bf 119} (2017) 181301},
  [\href{https://arxiv.org/abs/1705.06655}{{\tt 1705.06655}}].

\bibitem{Cui:2017nnn}
{\scshape PandaX-II} collaboration, X.~Cui et~al., \emph{{Dark Matter Results
  From 54-Ton-Day Exposure of PandaX-II Experiment}},
  \href{http://dx.doi.org/10.1103/PhysRevLett.119.181302}{\emph{Phys. Rev.
  Lett.} {\bf 119} (2017) 181302},
  [\href{https://arxiv.org/abs/1708.06917}{{\tt 1708.06917}}].

\bibitem{Fermi-LAT:2016uux}
{\scshape Fermi-LAT, DES} collaboration, A.~Albert et~al., \emph{{Searching for
  Dark Matter Annihilation in Recently Discovered Milky Way Satellites with
  Fermi-LAT}},
  \href{http://dx.doi.org/10.3847/1538-4357/834/2/110}{\emph{Astrophys. J.}
  {\bf 834} (2017) 110}, [\href{https://arxiv.org/abs/1611.03184}{{\tt
  1611.03184}}].

\bibitem{Ackermann:2015lka}
{\scshape Fermi-LAT} collaboration, M.~Ackermann et~al., \emph{{Updated search
  for spectral lines from Galactic dark matter interactions with pass 8 data
  from the Fermi Large Area Telescope}},
  \href{http://dx.doi.org/10.1103/PhysRevD.91.122002}{\emph{Phys. Rev.} {\bf
  D91} (2015) 122002}, [\href{https://arxiv.org/abs/1506.00013}{{\tt
  1506.00013}}].

\bibitem{Abramowski:2013ax}
{\scshape H.E.S.S.} collaboration, A.~Abramowski et~al., \emph{{Search for
  Photon-Linelike Signatures from Dark Matter Annihilations with H.E.S.S.}},
  \href{http://dx.doi.org/10.1103/PhysRevLett.110.041301}{\emph{Phys. Rev.
  Lett.} {\bf 110} (2013) 041301}, [\href{https://arxiv.org/abs/1301.1173}{{\tt
  1301.1173}}].

\bibitem{Garny:2013ama}
M.~Garny, A.~Ibarra, M.~Pato and S.~Vogl, \emph{{Internal bremsstrahlung
  signatures in light of direct dark matter searches}},
  \href{http://dx.doi.org/10.1088/1475-7516/2013/12/046}{\emph{JCAP} {\bf 1312}
  (2013) 046}, [\href{https://arxiv.org/abs/1306.6342}{{\tt 1306.6342}}].

\bibitem{Goodman:2010yf}
J.~Goodman, M.~Ibe, A.~Rajaraman, W.~Shepherd, T.~M.~P. Tait and H.-B. Yu,
  \emph{{Constraints on Light Majorana dark Matter from Colliders}},
  \href{http://dx.doi.org/10.1016/j.physletb.2010.11.009}{\emph{Phys. Lett.}
  {\bf B695} (2011) 185--188}, [\href{https://arxiv.org/abs/1005.1286}{{\tt
  1005.1286}}].

\bibitem{Athron:2017qdc}
{\scshape GAMBIT} collaboration, P.~Athron et~al., \emph{{Global fits of
  GUT-scale SUSY models with GAMBIT}},
  \href{http://dx.doi.org/10.1140/epjc/s10052-017-5167-0}{\emph{Eur. Phys. J.}
  {\bf C77} (2017) 824}, [\href{https://arxiv.org/abs/1705.07935}{{\tt
  1705.07935}}].

\bibitem{Randall:1999ee}
L.~Randall and R.~Sundrum, \emph{{A Large mass hierarchy from a small extra
  dimension}}, \href{http://dx.doi.org/10.1103/PhysRevLett.83.3370}{\emph{Phys.
  Rev. Lett.} {\bf 83} (1999) 3370--3373},
  [\href{https://arxiv.org/abs/hep-ph/9905221}{{\tt hep-ph/9905221}}].

\bibitem{Kraml:2017atm}
S.~Kraml, U.~Laa, K.~Mawatari and K.~Yamashita, \emph{{Simplified dark matter
  models with a spin-2 mediator at the LHC}},
  \href{http://dx.doi.org/10.1140/epjc/s10052-017-4871-0}{\emph{Eur. Phys. J.}
  {\bf C77} (2017) 326}, [\href{https://arxiv.org/abs/1701.07008}{{\tt
  1701.07008}}].

\bibitem{Goldberger:1999wh}
W.~D. Goldberger and M.~B. Wise, \emph{{Bulk fields in the Randall-Sundrum
  compactification scenario}},
  \href{http://dx.doi.org/10.1103/PhysRevD.60.107505}{\emph{Phys. Rev.} {\bf
  D60} (1999) 107505}, [\href{https://arxiv.org/abs/hep-ph/9907218}{{\tt
  hep-ph/9907218}}].

\bibitem{Davoudiasl:1999tf}
H.~Davoudiasl, J.~L. Hewett and T.~G. Rizzo, \emph{{Bulk gauge fields in the
  Randall-Sundrum model}},
  \href{http://dx.doi.org/10.1016/S0370-2693(99)01430-6}{\emph{Phys. Lett.}
  {\bf B473} (2000) 43--49}, [\href{https://arxiv.org/abs/hep-ph/9911262}{{\tt
  hep-ph/9911262}}].

\bibitem{Pomarol:1999ad}
A.~Pomarol, \emph{{Gauge bosons in a five-dimensional theory with localized
  gravity}}, \href{http://dx.doi.org/10.1016/S0370-2693(00)00737-1}{\emph{Phys.
  Lett.} {\bf B486} (2000) 153--157},
  [\href{https://arxiv.org/abs/hep-ph/9911294}{{\tt hep-ph/9911294}}].

\bibitem{Chang:1999nh}
S.~Chang, J.~Hisano, H.~Nakano, N.~Okada and M.~Yamaguchi, \emph{{Bulk standard
  model in the Randall-Sundrum background}},
  \href{http://dx.doi.org/10.1103/PhysRevD.62.084025}{\emph{Phys. Rev.} {\bf
  D62} (2000) 084025}, [\href{https://arxiv.org/abs/hep-ph/9912498}{{\tt
  hep-ph/9912498}}].

\bibitem{Davoudiasl:2000wi}
H.~Davoudiasl, J.~L. Hewett and T.~G. Rizzo, \emph{{Experimental probes of
  localized gravity: On and off the wall}},
  \href{http://dx.doi.org/10.1103/PhysRevD.63.075004}{\emph{Phys. Rev.} {\bf
  D63} (2001) 075004}, [\href{https://arxiv.org/abs/hep-ph/0006041}{{\tt
  hep-ph/0006041}}].

\bibitem{Geng:2018hpq}
C.-Q. Geng, D.~Huang and K.~Yamashita, \emph{{LHC Searches for Top-philic
  Kaluza-Klein Graviton}},
  \href{http://dx.doi.org/10.1007/JHEP10(2018)046}{\emph{JHEP} {\bf 10} (2018)
  046}, [\href{https://arxiv.org/abs/1807.09643}{{\tt 1807.09643}}].

\bibitem{Han:2015cty}
C.~Han, H.~M. Lee, M.~Park and V.~Sanz, \emph{{The diphoton resonance as a
  gravity mediator of dark matter}},
  \href{http://dx.doi.org/10.1016/j.physletb.2016.02.040}{\emph{Phys. Lett.}
  {\bf B755} (2016) 371--379}, [\href{https://arxiv.org/abs/1512.06376}{{\tt
  1512.06376}}].

\bibitem{Lee:2013bua}
H.~M. Lee, M.~Park and V.~Sanz, \emph{{Gravity-mediated (or Composite) Dark
  Matter}}, \href{http://dx.doi.org/10.1140/epjc/s10052-014-2715-8}{\emph{Eur.
  Phys. J.} {\bf C74} (2014) 2715},
  [\href{https://arxiv.org/abs/1306.4107}{{\tt 1306.4107}}].

\bibitem{Lee:2014caa}
H.~M. Lee, M.~Park and V.~Sanz, \emph{{Gravity-mediated (or Composite) Dark
  Matter Confronts Astrophysical Data}},
  \href{http://dx.doi.org/10.1007/JHEP05(2014)063}{\emph{JHEP} {\bf 05} (2014)
  063}, [\href{https://arxiv.org/abs/1401.5301}{{\tt 1401.5301}}].

\bibitem{Rueter:2017nbk}
T.~D. Rueter, T.~G. Rizzo and J.~L. Hewett, \emph{{Gravity-Mediated Dark Matter
  Annihilation in the Randall-Sundrum Model}},
  \href{http://dx.doi.org/10.1007/JHEP10(2017)094}{\emph{JHEP} {\bf 10} (2017)
  094}, [\href{https://arxiv.org/abs/1706.07540}{{\tt 1706.07540}}].

\bibitem{Geng:2016xin}
C.-Q. Geng and D.~Huang, \emph{{Note on spin-2 particle interpretation of the
  750 GeV diphoton excess}},
  \href{http://dx.doi.org/10.1103/PhysRevD.93.115032}{\emph{Phys. Rev.} {\bf
  D93} (2016) 115032}, [\href{https://arxiv.org/abs/1601.07385}{{\tt
  1601.07385}}].

\bibitem{Khojali:2016pvu}
M.~O. Khojali, A.~Goyal, M.~Kumar and A.~S. Cornell, \emph{{Minimal Spin-3/2
  Dark Matter in a simple $s$-channel model}},
  \href{http://dx.doi.org/10.1140/epjc/s10052-016-4589-4}{\emph{Eur. Phys. J.}
  {\bf C77} (2017) 25}, [\href{https://arxiv.org/abs/1608.08958}{{\tt
  1608.08958}}].

\bibitem{Khojali:2017tuv}
M.~O. Khojali, A.~Goyal, M.~Kumar and A.~S. Cornell, \emph{{Spin-3/2 Dark
  Matter in a simple $t$-channel model}},
  \href{http://dx.doi.org/10.1140/epjc/s10052-018-6407-7}{\emph{Eur. Phys. J.}
  {\bf C78} (2018) 920}, [\href{https://arxiv.org/abs/1705.05149}{{\tt
  1705.05149}}].

\bibitem{Creminelli:2001th}
P.~Creminelli, A.~Nicolis and R.~Rattazzi, \emph{{Holography and the
  electroweak phase transition}},
  \href{http://dx.doi.org/10.1088/1126-6708/2002/03/051}{\emph{JHEP} {\bf 03}
  (2002) 051}, [\href{https://arxiv.org/abs/hep-th/0107141}{{\tt
  hep-th/0107141}}].

\bibitem{Csaki:1999mp}
C.~Csaki, M.~Graesser, L.~Randall and J.~Terning, \emph{{Cosmology of brane
  models with radion stabilization}},
  \href{http://dx.doi.org/10.1103/PhysRevD.62.045015}{\emph{Phys. Rev.} {\bf
  D62} (2000) 045015}, [\href{https://arxiv.org/abs/hep-ph/9911406}{{\tt
  hep-ph/9911406}}].

\bibitem{Semenov:2014rea}
A.~Semenov, \emph{{LanHEP — A package for automatic generation of Feynman
  rules from the Lagrangian. Version 3.2}},
  \href{http://dx.doi.org/10.1016/j.cpc.2016.01.003}{\emph{Comput. Phys.
  Commun.} {\bf 201} (2016) 167--170},
  [\href{https://arxiv.org/abs/1412.5016}{{\tt 1412.5016}}].

\bibitem{Belyaev:2012qa}
A.~Belyaev, N.~D. Christensen and A.~Pukhov, \emph{{CalcHEP 3.4 for collider
  physics within and beyond the Standard Model}},
  \href{http://dx.doi.org/10.1016/j.cpc.2013.01.014}{\emph{Comput. Phys.
  Commun.} {\bf 184} (2013) 1729--1769},
  [\href{https://arxiv.org/abs/1207.6082}{{\tt 1207.6082}}].

\bibitem{Alarcon:2011zs}
J.~M. Alarcon, J.~Martin~Camalich and J.~A. Oller, \emph{{The chiral
  representation of the $\pi N$ scattering amplitude and the pion-nucleon sigma
  term}}, \href{http://dx.doi.org/10.1103/PhysRevD.85.051503}{\emph{Phys. Rev.}
  {\bf D85} (2012) 051503}, [\href{https://arxiv.org/abs/1110.3797}{{\tt
  1110.3797}}].

\bibitem{Hisano:2015bma}
J.~Hisano, R.~Nagai and N.~Nagata, \emph{{Effective Theories for Dark Matter
  Nucleon Scattering}},
  \href{http://dx.doi.org/10.1007/JHEP05(2015)037}{\emph{JHEP} {\bf 05} (2015)
  037}, [\href{https://arxiv.org/abs/1502.02244}{{\tt 1502.02244}}].

\bibitem{Kopp:2014tsa}
J.~Kopp, L.~Michaels and J.~Smirnov, \emph{{Loopy Constraints on Leptophilic
  Dark Matter and Internal Bremsstrahlung}},
  \href{http://dx.doi.org/10.1088/1475-7516/2014/04/022}{\emph{JCAP} {\bf 1404}
  (2014) 022}, [\href{https://arxiv.org/abs/1401.6457}{{\tt 1401.6457}}].

\bibitem{Kopp:2009et}
J.~Kopp, V.~Niro, T.~Schwetz and J.~Zupan, \emph{{DAMA/LIBRA and leptonically
  interacting Dark Matter}},
  \href{http://dx.doi.org/10.1103/PhysRevD.80.083502}{\emph{Phys. Rev.} {\bf
  D80} (2009) 083502}, [\href{https://arxiv.org/abs/0907.3159}{{\tt
  0907.3159}}].

\bibitem{DEramo:2017zqw}
F.~D'Eramo, B.~J. Kavanagh and P.~Panci, \emph{{Probing Leptophilic Dark
  Sectors with Hadronic Processes}},
  \href{http://dx.doi.org/10.1016/j.physletb.2017.05.063}{\emph{Phys. Lett.}
  {\bf B771} (2017) 339--348}, [\href{https://arxiv.org/abs/1702.00016}{{\tt
  1702.00016}}].

\bibitem{Aprile:2015uzo}
{\scshape XENON} collaboration, E.~Aprile et~al., \emph{{Physics reach of the
  XENON1T dark matter experiment}},
  \href{http://dx.doi.org/10.1088/1475-7516/2016/04/027}{\emph{JCAP} {\bf 1604}
  (2016) 027}, [\href{https://arxiv.org/abs/1512.07501}{{\tt 1512.07501}}].

\bibitem{Aprile:2017aty}
{\scshape XENON} collaboration, E.~Aprile et~al., \emph{{The XENON1T Dark
  Matter Experiment}},
  \href{http://dx.doi.org/10.1140/epjc/s10052-017-5326-3}{\emph{Eur. Phys. J.}
  {\bf C77} (2017) 881}, [\href{https://arxiv.org/abs/1708.07051}{{\tt
  1708.07051}}].

\bibitem{Abdalla:2016olq}
{\scshape H.E.S.S.} collaboration, H.~Abdalla et~al., \emph{{H.E.S.S. Limits on
  Linelike Dark Matter Signatures in the 100 GeV to 2 TeV Energy Range Close to
  the Galactic Center}},
  \href{http://dx.doi.org/10.1103/PhysRevLett.117.151302}{\emph{Phys. Rev.
  Lett.} {\bf 117} (2016) 151302},
  [\href{https://arxiv.org/abs/1609.08091}{{\tt 1609.08091}}].

\bibitem{Ackermann:2015zua}
{\scshape Fermi-LAT} collaboration, M.~Ackermann et~al., \emph{{Searching for
  Dark Matter Annihilation from Milky Way Dwarf Spheroidal Galaxies with Six
  Years of Fermi Large Area Telescope Data}},
  \href{http://dx.doi.org/10.1103/PhysRevLett.115.231301}{\emph{Phys. Rev.
  Lett.} {\bf 115} (2015) 231301},
  [\href{https://arxiv.org/abs/1503.02641}{{\tt 1503.02641}}].

\bibitem{Sirunyan:2018ryr}
{\scshape CMS} collaboration, A.~M. Sirunyan et~al., \emph{{Search for resonant
  $ \mathrm{t}\overline{\mathrm{t}} $ production in proton-proton collisions at
  $ \sqrt{s}=13 $ TeV}},
  \href{http://dx.doi.org/10.1007/JHEP04(2019)031}{\emph{JHEP} {\bf 04} (2019)
  031}, [\href{https://arxiv.org/abs/1810.05905}{{\tt 1810.05905}}].

\bibitem{Bonciani:2015hgv}
R.~Bonciani, T.~Jezo, M.~Klasen, F.~Lyonnet and I.~Schienbein,
  \emph{{Electroweak top-quark pair production at the LHC with $Z'$ bosons to
  NLO QCD in POWHEG}},
  \href{http://dx.doi.org/10.1007/JHEP02(2016)141}{\emph{JHEP} {\bf 02} (2016)
  141}, [\href{https://arxiv.org/abs/1511.08185}{{\tt 1511.08185}}].

\end{thebibliography}\endgroup
 \bibliographystyle{JHEP}
\end{document}